\documentclass[11pt]{iopart}

\usepackage{iopams}

\usepackage{graphicx}
\usepackage{hyperref}
\usepackage{float}
\usepackage{xcolor}
\usepackage[justification=justified,font=footnotesize,format=plain,labelfont=bf,labelsep=period]{caption}
\usepackage{multirow}
\usepackage{booktabs}
\usepackage{stmaryrd}

\expandafter\let\csname equation*\endcsname\relax
\expandafter\let\csname endequation*\endcsname\relax
\usepackage{amsmath}

\usepackage[backend=biber,style=authoryear,maxnames=2]{biblatex}
\begin{document}

\title[Machine Learning of Mesoscale Fluxes in Midlatitudes]{Machine Learning of Vertical Fluxes by Unresolved Midlatitude Mesoscale Processes}

\author{Erisa Ismaili$^{1}$; Robert C. Jnglin Wills$^{1}$; Tom Beucler$^{2}$}
\address{$^{1}$ETH Zürich, $^{2}$University of Lausanne}
\ead{r.jnglinwills@usys.ethz.ch} 
\vspace{10pt}

\begin{abstract}

\noindent 
Machine learning (ML) can represent processes unresolved in coarse-resolution Earth system models (ESMs) by learning from high-resolution climate data. Such ML parameterization approaches have been primarily tested in idealized setups (e.g., aquaplanets), where they have focused on deep convection. It remains largely unexplored whether these approaches could be used in a more targeted fashion to learn vertical fluxes resulting from midlatitude mesoscale processes, such as slantwise convection and frontal dynamics in extratropical cyclones, which are not well represented in ESMs. 
\
To address this, we employ a variable-resolution CESM2 simulation with a refined area over the North Atlantic (14-km grid refinement) that resolves such midlatitude mesoscale processes. We train an artificial neural network to predict vertical profiles of mesoscale moisture, heat, and momentum fluxes from the perspective of a coarse-resolution (111-km grid) model. We then assess feature importance and the vertical localization relationship between inputs and outputs through ablation experiments and explainable artificial intelligence.
\
Our results show that a large number of features are required to achieve reasonable model performance when data come from the midlatitudes of real-geography atmospheric simulations, especially when coarse-grained vertical velocities, which we show are not representative of vertical velocities in a coarse-resolution model, are excluded as inputs. Feature importance analysis reveals the importance of vertically non-local information in temperature, moisture, and the meridional wind. Information from neighboring columns also helps to predict mesoscale fluxes, e.g., through the action of conditional symmetric instability. We suggest that these non-local relationships capture the influence of cold air outbreaks and fronts on mesoscale fluxes.  
\
Our results demonstrate the importance of vertically non-local processes, clarify the regime-dependent predictability of mesoscale fluxes, and identify variables most informative for their parameterization, providing guidance for improving ESMs with ML and advancing our understanding of multi-scale interactions in the midlatitudes.




\end{abstract}

%
%
%
%
%

\section{Introduction}
\label{sec:introduction}


Together with the help of higher-resolution simulations, improved process knowledge, and validation with new observational data, machine learning (ML) has the potential to improve Earth System Models (\cite{schneider_opinion_2024, irrgang_towards_2021, eyring_pushing_2024,Mansfield2023}).
This includes advancing climate models by developing and implementing ML parameterizations, which represent small-scale processes at reduced computational cost by establishing relationships between atmospheric state variables and parameterized process outputs, such as fluxes or tendencies, through data-driven models (e.g., \cite{Gentine_2018, Brenowitz2019, Yuval_OGorman_2021}). 
Despite their promise to empower next-generation climate models (\cite{eyring2024ai,Beucler2024_SATW_HybridESM}), ML models do not provide explanations for their predictions unless they are complemented by causal inference or explainable AI (XAI) approaches that help understand the patterns learned by ML (\cite{irrgang_towards_2021}). 
 
Previous studies have successfully applied ML techniques for parameterizations of subgrid-scale processes that lead to stable and accurate simulations (e.g., \cite{Rasp2018,Lin2025}). These efforts have primarily focused on the parameterization of deep convection. Midlatitude mesoscale processes such as slantwise convection and frontal dynamics are not explicitly treated in such approaches, are only partially represented in coarse-resolution Earth system models (ESMs), and remain largely unrepresented in conventional parameterizations. As the ML parameterization community shifts from idealized geometries (e.g., aquaplanets, \cite{Wang_nonlocal,yuval_neural-network_2023}) to more realistic configurations for comprehensive climate modeling (\cite{watts2024NNrealisticgeography,heuer2025trainingdataconfidenceguidedmixing}), the need for process-oriented evaluation grows, particularly because regional and process-specific biases persist even when global performance metrics are competitive (\cite{yu2023climsim}). To address this gap, we use ML to model and understand midlatitude mesoscale processes in a comprehensive climate model (CESM2), focusing specifically on the Gulf Stream region.
%

Mesoscale atmospheric processes that are neither resolved nor parameterized in global climate models (i.e., those at scales below the $\sim$500 km effective resolution of a typical 100-km ESM; \cite{klaver}) can have an important upscale impact on climate variability and change. An example of such mesoscale influence on the climate is demonstrated by \cite{wills_resolving_2024}, who show that circulation responses to surface anomalies in the Gulf Stream region are increased when heat and momentum fluxes by mesoscale processes are resolved. Oceanic western boundary currents such as the Gulf Stream transfer heat to the atmosphere, especially during the cold season. Surface sensible and latent heat fluxes are further enhanced during cold air outbreaks (\cite{OceanMesoscale_2023, CAO_2015}). Sea surface temperature (SST) fluctuations in these regions imprint on local climate variability, for example as seen in vertical motion and precipitation (\cite{larson_signature_2024, AtmosphericResponse_2010}). Vertical fluxes of momentum, heat, and moisture are important for communicating anomalies from the boundary layer into the free troposphere (\cite{wills_resolving_2024}). At the mesoscale, these fluxes occur primarily due to convection, including slantwise convection, and frontal processes. However, vertical fluxes can also arise due to orography, which induces atmospheric circulation responses across a range of scales including the mesoscale (\cite{LandscapeInduced, Orographic_resolution}), as well as from gravity waves triggered by orography, convection, jets, or fronts. Gravity waves are especially relevant for momentum fluxes (\cite{gupta_gravity_2024, Achatz_gravity_2024}), and have a more limited contribution to heat fluxes (\cite{GW_heatflux_2021}). 

Over the Gulf Stream and in general near western boundary currents, there is evidence that slantwise convection occurs with an enhanced frequency in winter (\cite{Czaja_Blunt_2011, Korty_Schneider_2007, Sheldon_Czaja_2014}).
Slantwise convection typically appears as an elongated, band-like structure, and while the convective motion is large enough to be resolved by the coarse grid spacing, the underlying conditional symmetric instability (CSI) is subgrid-scale and requires parameterization in global models (\cite{WRFModelParameterization_2022}).
CSI is a mechanism that reduces the vertical shear of horizontal wind and is identifiable by the presence of downward mesoscale momentum fluxes.
However, CSI release does not always occur spontaneously. Frontogenesis can help establish conditions for slantwise convection in situations where small perturbations might not otherwise be sufficient to develop slantwise convection (\cite{book_mesoscaleMeteo_2010}). 
Slantwise convection, in particular, is known to lead to banded precipitation that appears often in the Gulf Stream region within extratropical cyclones (\cite{seltzer_possible_1985}), although \cite{Misuse_CSI_1999} have noted the mis- and overuse of slantwise convection as a sole explanation for banded precipitation in the midlatitudes and highlighted the close interplay of slantwise convection and frontogenesis.

In this study, we develop and analyze a targeted ML parameterization of the mesoscale processes leading to subgrid-scale vertical fluxes of momentum, heat, and moisture in midlatitudes. We use novel front-resolving variable-resolution CESM2.2 simulations from \cite{wills_resolving_2024}, which have resolution refined to 14 km over the North Atlantic. This horizontal grid spacing is able to effectively resolve weather systems larger than roughly 70 km (\cite{klaver}) and thus is able to specifically target mesoscale processes important in midlatitude baroclinic zones, such as slantwise convection, while smaller-scale processes such as deep convection continue to be handled by traditional parameterizations. Building on prior ML-parameterization studies showing that ANNs can effectively learn subgrid relationships from high-resolution simulations (\cite{ML_clouds_climate_2023}), we train simple feedforward ANNs to predict subgrid-scale flux profiles.
The targets of the ANN are the vertical profiles of mesoscale fluxes, and the input features consist of vertical profiles of atmospheric state variables (temperature, momentum, and moisture) and neighboring profile information to compute the horizontal derivatives of a selection of variables. In midlatitudes, neighboring profile information has been shown to add skill in the ML parameterization work by \cite{Wang_nonlocal}. This is especially of interest when processes like slantwise convection follow tilted motion.
To gain intuition about how the ANN predicts mesoscale fluxes, we assess which features matter for predicting the fluxes. We first use ablation experiments, in which we remove subsets of features and retrain the ANN; this is computationally expensive but provides a robust measure of each feature's overall informativeness. We then complement this with SHAP, a game-theory-based XAI method that is cheaper to apply and also reveals the functional role and sign of each feature's contribution to the predictions. Together, these analyses clarify links between mesoscale fluxes and atmospheric state variables. 


The structure of the paper is as follows. Section \ref{sec:methods} introduces the climate data used in the study, the exact definition of the features and targets of the ANN, as well as the pre-processing that is applied to the data before it is passed into the ML model. Section \ref{sec:ML} presents the ML methods, including the neural network architecture and feature importance analysis. In the results section (Section \ref{sec:results}), we start by defining regimes, which give a first sense of the processes determining mesoscale fluxes in the study region and will be used to assess the generalizability of the ML model to various atmospheric situations. We then present the results of the ANN prediction together with a comparison of the performance across different regimes. We also present the results of the ablation experiments and answer the questions of which features add skill to the prediction of vertical mesoscale fluxes. We continue with an analysis of the vertical localization relationship of inputs and outputs. Finally, we extract physical insights from the SHAP feature importance analysis. Section \ref{sec:discusssion} provides discussion and conclusions based on our results.


\section{Data}
\label{sec:methods}
\subsection{Climate Model Data for Training}
\label{subsec:Data}
The dataset used to train the ANN is constructed from the outputs of a simulation by \cite{wills_resolving_2024}, conducted with the Community Atmospheric Model version 6 (CAM6) within the Community Earth System Model version 2.2 (CESM2.2). CESM offers the option to run the simulations with the spectral element (SE) method which allows for regional refinements as described in \cite{lauritzen_ncar_2018}. In our case, the regionally refined grids are done on the basis of the quasi-uniform ne30pg3 configuration. On this grid, each cubed-sphere face has 30 $\times$ 30 quadrilateral elements and each element is divided in $3 \times 3$ control volumes. This results in an average horizontal grid spacing of 110 km. To get a regional refinement over the North Atlantic, the grid is rotated such that a cubed-sphere face lies in the center of the North Atlantic. Refinement by a factor of 8 is used over the target region, resulting in a regional refinement to a horizontal grid spacing of 14 km in the North Atlantic. The effective resolution is at least 5 times higher; thus the model is able to represent weather systems larger than roughly 70 km (\cite{klaver}). This simulation provides the high resolution information on which the ML model is trained. We will refer to the simulation on this grid with a refinement factor of 8 over the North Atlantic as the NATLx8 simulation following the original study. CAM6 uses conventional (i.e., non-ML) parameterization schemes for subgrid-scale processes such as turbulence, cloud microphysics, shallow convection, deep convection, surface fluxes, and orographic gravity waves. The data used in this study correspond to the NATLx8 reference simulation in \cite{wills_resolving_2024}, which is performed using year-2000 radiative forcing and specified seasonally varying SST and sea-ice climatology. For further information on the simulation setup, refer to \cite{wills_resolving_2024}.

\begin{figure}[h]
    \centering
    \includegraphics[width=1\linewidth]{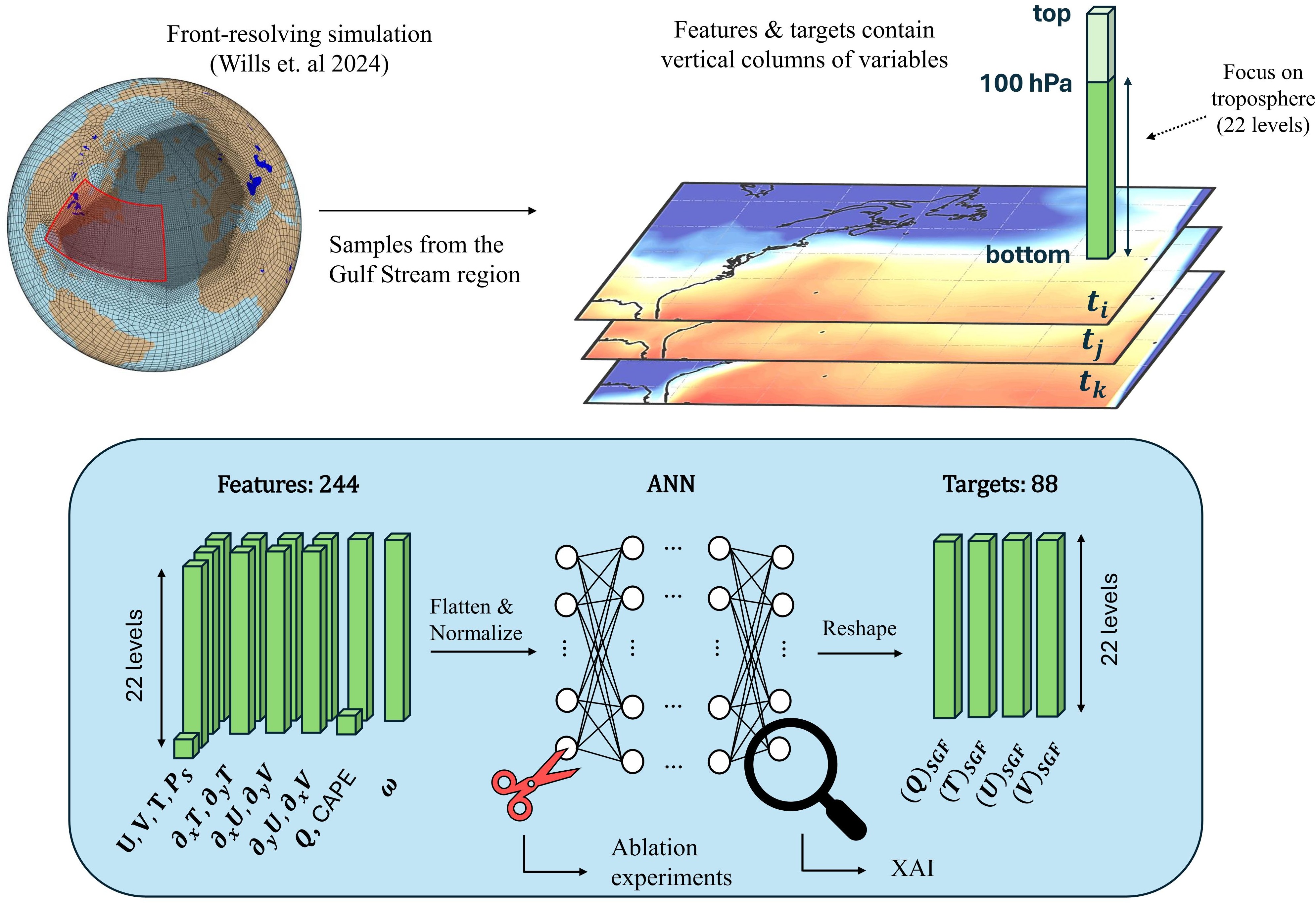}
    \caption{Schematic explaining the methods for training an artificial neural network (ANN) to learn mesoscale atmospheric fluxes. A front-resolving simulation by \cite{wills_resolving_2024} provides the data with a refined resolution over the North Atlantic. We focus on the Gulf Stream region (red area), where columns of atmospheric variables are used to construct the dataset containing the features and targets to train the ANN. To focus on the troposphere, only 22 out of 32 levels are used. The 244 features contain the profile information of 11 atmospheric state variables and single-level information of the surface pressure and CAPE variables. The 88 targets contain the columns of the subgrid-scale fluxes of moisture, heat, and the horizontal momenta. Features and target variables are coarse-grained to a resolution of roughly 100 km. Using entire profile information for the feature and target variables is referred to as profile-to-profile mapping. Ablation experiments and XAI methods are used to assess feature importance.}
    \label{fig:ANN_schematic_v2}
\end{figure}

From the simulation output, we use a subset of the data in space. The simulation output has 32 hybrid pressure-sigma levels representing the vertical dimension, with a model top at 2 hPa. For this work, we select the lowest 22 of the 32 levels, aiming to mainly represent the tropospheric convective processes up to 100 hPa. This is motivated by the relatively poor performance of neural networks at levels higher than approximately 100 hPa in previous works (\cite{Wang_nonlocal}), as was also seen in preliminary tests in this work (not shown). 
Horizontally, we focus on the Gulf Stream region and select
the longitudes 90$^{\circ}$W-30$^{\circ}$W and the latitudes 25$^{\circ}$N-55$^{\circ}$N (red area on the grid in Figure \ref{fig:ANN_schematic_v2}). 

\subsection{Selection of the Features and Targets}
\label{subsec:definition of inputs and outputs}

The goal of the ANN is to predict the vertical columns of the vertical subgrid-scale fluxes per grid point in the selected Gulf Stream region. The features of the ANN characterize the coarse-scale environment, including column variables with vertical profile information (e.g., temperature) and scalar variables with only one value per sample (e.g., surface pressure). To reduce under-determination and limit the influence of static geographic proxies, we exclude prescribed SST and land fraction as predictors, which encourages the ANN to base its flux predictions more strongly on the vertical profiles of temperature, humidity, and winds. A schematic illustrating the process of feeding inputs into the ANN and generating the mesoscale flux outputs is shown in Figure \ref{fig:ANN_schematic_v2}. The input and output variables are within the high-resolution region and are coarse-grained to mimic a typical coarser-resolution climate model. To follow best practices in climate modeling, this is achieved by using the NetCDF Operators (NCO) function \texttt{ncremap} (\cite{conservative_remapping}) to remap the variables conservatively onto a 1.25$^{\circ}$ longitude $\times$ 0.94$^{\circ}$ latitude grid, corresponding to a horizontal resolution of approximately 100 km in the midlatitudes.

\begin{figure}[h]
    \centering
    \includegraphics[width=1\linewidth]{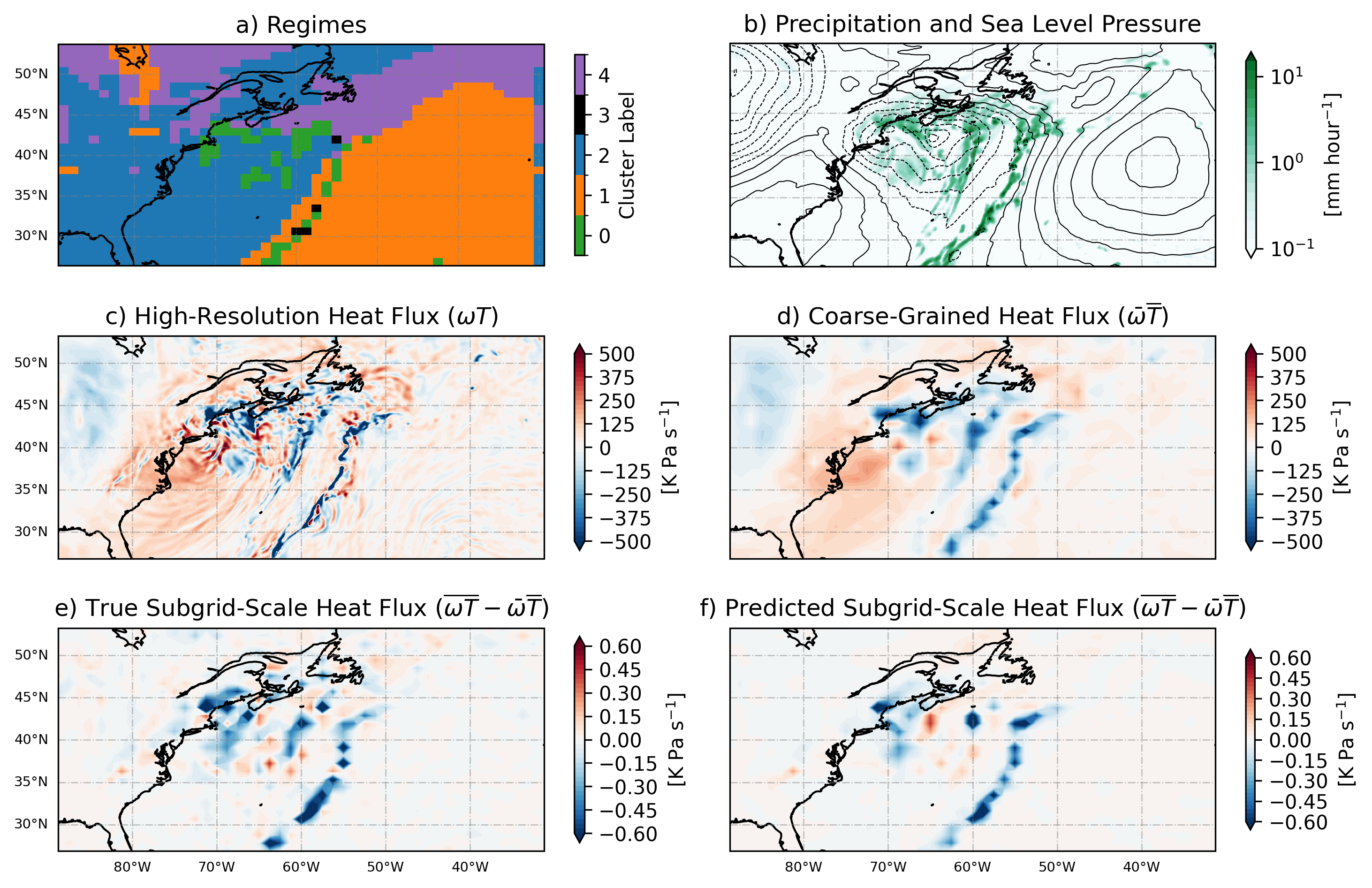}
    \caption{Example event of an extratropical cyclone at model time stamp 0032-12-12 00:00:00 with a map of a) categorical regimes derived from a k-means clustering with coarse-grained variables (see Section \ref{subsec:regime def}), b) the high-resolution total precipitation and the sea level pressure anomalies, c) the high-resolution heat flux, d) the coarse-grained heat flux, e) the true subgrid-scale flux and f) the predicted subgrid-scale flux by the ANN. The selected level for the heat flux is the nearest level to 700 hPa.}
    \label{fig:intro_plot}
\end{figure}

We focus on predicting four subgrid-scale vertical fluxes: those of zonal wind $U$, meridional wind $V$, temperature $T$, and moisture $Q$. The subgrid-scale vertical fluxes are computed using the prognostic variable that is transported vertically $\phi \in \{U, V, T, Q\}$ along with the vertical velocity $\omega$. Using Reynolds decomposition the following expression determines the subgrid-scale flux $(\phi)_{\text{SGF}}$ as the covariance between $\phi$ and $\omega$:
\begin{equation}
    (\phi)_{\text{SGF}} \equiv \overline{\phi' \omega'} =\overline{\omega \phi} - \overline{\omega}\ \overline{\phi}.
    \label{eq:SGF_DEF}
\end{equation}
Here, the overline denotes coarse-graining of the high-resolution field via conservative remapping with \texttt{ncremap} and the primes represent the deviation from the coarse-grained field. Equation~\ref{eq:SGF_DEF} defines subgrid-scale fluxes as the fluxes remaining after subtracting the low-resolution fluxes (where variables are coarse-grained before multiplication) from the coarse-grained high-resolution fluxes (where coarse-graining is applied after multiplication). Figures \ref{fig:intro_plot}c and \ref{fig:intro_plot}d highlight the difference between the high-resolution heat flux $\omega T$ and the heat flux $\overline{\omega}\ \overline{T}$ computed from coarse-grained variables based on an example event. Example subgrid fluxes for the same event are shown in Figure \ref{fig:intro_plot}e. Note that while the subgrid vertical heat fluxes are small compared to the total vertical heat fluxes in a snapshot, they represent a non-negligible contribution to the vertical heat flux climatology in this region (\cite{wills_resolving_2024}). Our choice of output variables results in 88 entries for the output layer after flattening the output flux columns to a single output vector. In this study, the horizontal fluxes are neglected as they are expected to be less relevant (e.g., compared to fluxes by large-scale motions) when focusing on convective processes and vertical mixing of prognostic variables. Since slantwise convection results in tilted motion, including horizontal fluxes in the output vector may nevertheless be worth considering in future work.

The ANN inputs consist of vertical profiles of atmospheric state variables $T$, $U$, $V$, $\omega$, and $Q$, together with scalar surface pressure $P_s$ and convective available potential energy (CAPE). To incorporate information from neighboring grid cells, we add centered-difference horizontal derivatives $\partial_x$ and $\partial_y$ of $U$, $V$, and $T$. This differs from the non-local parameterization in \cite{Wang_nonlocal}, which uses $3\times3$ neighboring columns for each profile variable. In total, our setup includes 11 profile variables and 2 scalar variables, yielding 244 input features; much less than using the $3\times3$ column approach. We would also have included sensible and latent surface heat fluxes, given their relevance for convection, but these fields are not available in the 6-hourly NATLx8 outputs.

\subsection{Sampling Strategy and Data Partitioning}

The NATLx8 simulation data contains a total of 30 years with four time steps per day (6-hourly data).
To capture the diurnal and seasonal atmospheric cycles, we subsample the data by selecting the first 5 days (20 time steps) every 30 days. Over the 30-year span of the simulations and including all coarse-grained grid cell in the Gulf Stream region, this results in a dataset of nearly 10 million samples for training the ANN. Increasing the sample size to include 15 days every 30 days did not improve the performance of the ANN, suggesting that the chosen sampling approach already leads to convergence.

The data is divided into training, validation, and test sets. The training set includes the first 80\% of the time range, the validation set covers the next 10\%, and the test set contains the final 10\% of the time range. 
We construct an additional dataset within the time range of the test set by including the remaining 25 days of each 30-day period that were not initially selected, making it 6 times larger than the original test set. From this dataset, new test sets are generated by randomly drawing 5-day chunks, keeping a structure similar to the original test set. This allows for uncertainty estimation by bootstrapping, where we test the ML models on 1000 new datasets to get a spread of test performance. 

\subsection{Data Pre-Processing and Normalization}

Before training the ANN, the data is pre-processed to optimize the models' performance. 
The subgrid fluxes are heavy-tailed, reaching values to the order of 100 standard deviations when being standardized for each feature separately. These heavy tails are asymmetric with different skewness for each level and variable, and highly influence how the model will learn. It is desirable to accurately learn the tail values as they contain the interesting subgrid-scale information, as opposed to values close to zero, but one has to be careful of how they influence the learning of the model. 

We tried different normalization options. The input variables can be standardized for each vertical level separately or across the entire profile of the variable. Standardization is the easiest way to ensure unit consistency in all input features. In our case, the level-wise standardization option for the inputs resulted in better predictions. For the outputs, a simple normalization worked well: Each variable is divided by some constant to ensure similar value ranges and, therefore, maintain balanced weighting during training and avoid biased representation. Consequently, their units are arbitrary until they are transformed back. The constants for the normalization of outputs are chosen $c_Q = 3.33\times10^{-5}$ Pa s$^{-1}$, $c_T = 2.77\times10^{-2}$ K Pa s$^{-1}$, $c_U=c_V=0.1$ m Pa s$^{-2}$,
where the subscripts indicate the variable which is transported by the subgrid-scale flux (i.e., $c_U$ is the constant by which all levels of the $(U)_{\text{SGF}}$ column are normalized). 

\section{Machine Learning Methodology}
\label{sec:ML}

\subsection{Neural Network Architecture}
\label{subsec:ANN description}


We distinguish between fixed hyperparameters and parameters explored during model optimization.  A detailed overview of the hyperparameters is provided in Table \ref{tab: hyperparameters} of the Supplementary Information. The hyperparameters explored were tuned by first performing a coarse \texttt{optuna} search and then cross-validating for refinement.
We combine the training and validation sets to select a suitable ANN architecture and optimize the hyperparameters. Cross-validation is performed by splitting this combination into five folds and training the model five times. The architecture and hyperparameters of the neural network are optimized using the package \texttt{optuna} for an increased efficiency by pruning poorly performing trials in advance (\cite{optuna}). To mitigate the negative impact of the output variables' heavy-tailed distributions on validation error, we explore loss functions that grow linearly with the training error for outliers. Thus, the options for the loss function include the mean absolute error (MAE, L1 loss), Huber loss and the balanced L1 loss introduced by \cite{pang2019librarcnnbalancedlearning}.
This balanced L1 loss leads to the lowest validation MSE. The final neural network architecture consists of 8 hidden layers of a large width, batch normalization layers, and the GELU activation function. The exact structure is presented in Table~\ref{tab: network layers} in Supporting Information. When using the smooth GELU activation function, we do not include dropout layers as a regularization technique, as this activation function aims to combine the effects that the deterministic ReLU function and the stochastic dropout technique have on the neurons (\cite{GELU_paper}). The weights and biases are updated using the Adam optimizer with a weight decay of 4.07$\times10^{-6}$, where a cycling learning rate scheduler for the learning rate is used, inspired by the work of \cite{Wang_nonlocal}. The hyperparameters of the scheduler are the base learning rate of 2.15$\times10^{-6}$ and a maximum learning rate of 2.78$\times10^{-4}$. The cyclic mode is "triangular2" and step of the cycle is chosen such that after 4 epochs the learning rate returns to the base learning rate. We use early stopping and identify the minimum loss on the validation set at epoch 16. The training and test loss curves are shown in Figure \ref{fig: loss functions} in Supporting Information. 

\subsection{Feature Importance and Localization Experiments}
\label{sec:Fatures importance assessment description}

Ablation experiments can help assess the ability to successfully predict the targets with only a subset of the available features. Since the features consist of different variables at different levels, the ablation experiments can be used to assess the importance of certain variables as features, or to assess the vertical localization relationship. To investigate the impact of a more localized mapping on the ANN performance, we replace the profile-to-profile mapping, where vertical profiles of input variables are used to predict vertical profiles of output variables, with a level-to-level mapping, where each level of the vertical profile is predicted individually using the input variables on the same level. It is expected that this change in the localization relationship reduces the performance, since the direct dependence on information from other levels within the vertical profile is removed. We extend this setup to understand the extent of improvement when feature variable information is included from the levels directly above and below the target level. This means that each target level is now predicted using the corresponding input level as well as its nearest vertical neighbors, and we call this mapping the neighbor-to-level mapping. The level-to-level and neighbor-to-level mapping have a reduced number of ML model parameters by dividing the size of each hidden layer of the ANN by a factor of 8. This reduction of the width of the ANN is necessary to account for the significantly smaller number of features and targets. 

\subsection{Shapley Analysis for Physical Interpretability}
SHAP (SHapley Additive exPlanations) is an XAI method based on game theory that helps interpret predictions from complex models such as neural networks, which are intrinsically difficult to understand (\cite{shap_paper}). SHAP computes Shapley values that quantify how much each input feature contributes to a given output for a fixed trained model (\cite{shapley1953value}). This differs from ablation experiments, which assess the informativeness of a feature set for a model class by removing features and retraining: ablation measures how performance changes when the learning problem is re-solved without certain inputs, whereas SHAP attributes predictions within the already trained network. 

While typically cheaper than full ablation studies, SHAP can still be extremely expensive on large datasets, motivating its computation on a representative subset of samples. This introduces sampling uncertainty and requires sensitivity tests. We therefore choose enough subsets to ensure that at least one sample from each regime defined in Section~\ref{subsec:regime def} is included. We feed 2000 background samples into SHAP's DeepExplainer (the explainer designed for neural networks) to compute Shapley values for a randomly selected batch of 1000 samples, which we call the SHAP test data. To test robustness, we repeat this procedure for 10 subsets with different samples chosen randomly.

Summing the resulting Shapley values $\text{SHAP}(x_i, y_j)$ over the input features $x_i$ for a given output $y_j$ equals the difference $y_j^{\prime}$ between the ML model's prediction and the training-set average estimated using background samples (\cite{shap_paper}):
\begin{equation}
    \sum_{i}\text{SHAP}(x_i, y_j) = y_j^{\prime}.
\end{equation}
The sign of the Shapley values indicates whether, relative to this baseline, an input feature pushes the output to larger ($\text{SHAP}>0$) or smaller ($\text{SHAP}<0$) values for a given sample.
The sign can be ignored if we are only interested in the overall magnitude of the influence of the features. This magnitude is estimated by taking the average of the absolute Shapley values across samples. The resulting magnitudes can help, for example, understand what crucial information is missing if there is a significant gap between the neighbor-to-level mapping and the profile-to-profile mapping. The sign information is also relevant, as it may provide physical insight by indicating the response of the outputs when perturbing features in a given direction.

\cite{Shap_Beucler} introduced a nonlinear feature matrix $M$ to assess whether input deviations align, on average, with the sign of the Shapley values:
\begin{equation}
    M_{ij} = \left\langle \text{sign}(x_i^{\prime}) \times \text{SHAP}(x_i, y_j)\right\rangle,
\end{equation}
where, analogously to $y_j^{\prime}$, $x_i^{\prime}$ is the deviation of feature $x_i$ from its training-set mean. The expectation brackets $\langle \cdot \rangle$ denote an average over the SHAP test data.


\section{Results}
\label{sec:results}

\subsection{Regime Definition via K-means Clustering}
\label{subsec:regime def}

To gain a first intuition of the data, we define regimes using the K-means++ algorithm applied to a subset of our input and output variables. This analysis will also help interpret the ANN, by investigating the performance and the XAI methods on the different clusters, revealing the generalization ability of the ML models. The resulting clusters may not have a clear physical interpretation, but defining them helps distinguish different regimes relevant to subgrid parameterization. There is no correct number of clusters according to an elbow plot, since it does not show any apparent elbow (not shown), and creating a silhouette plot for the sample size of our dataset is computationally intractable. To maintain an overview and simultaneously have enough variability across regimes, we choose  $k=$ 5 clusters. We chose the feature space to be spanned by the following 12 variables: $U_{500}-U_{850}$ and $V_{500}-V_{850}$ are the difference of zonal/meridional wind at levels closest to 500 hPa and 850 hPa, and are a rough estimate of the vertical wind shear. We include the divergence $D_{\text{bot}} = (\partial_xU + \partial_yV) \big|_{\text{bot}}$ at the lowest level, and the relative vorticity $\zeta_{500} =  (\partial_xV - \partial_yU) \big|_{\text{500}}$ at the level nearest to 500 hPa, as these level choices helped in the formation of distinct clusters. Additionally, we include $T_{\text{bot}}-T_{850}$,  the difference in temperature between the bottom level and at the level closest to 850 hPa. The bottom level temperature of the atmosphere is used as a substitute for the skin temperature $T_{\text{skin}}$, needed to identify cold air outbreaks (\cite{CAO_2015}). Clearer and more distinct clusters are formed using the relative humidity RH instead of the specific humidity $Q$. The level choice for the humidity feature is the level closest to the surface. The remaining features are surface pressure, CAPE, and the four output subgrid-scale flux variables at the level closest to 500 hPa. A list of the features and of additional diagnostic quantities important in our cluster analysis can be found in Table \ref{tab: cluster features and vars} in the supporting information.

\begin{figure}[h]
    \centering
    \includegraphics[width=1\linewidth]{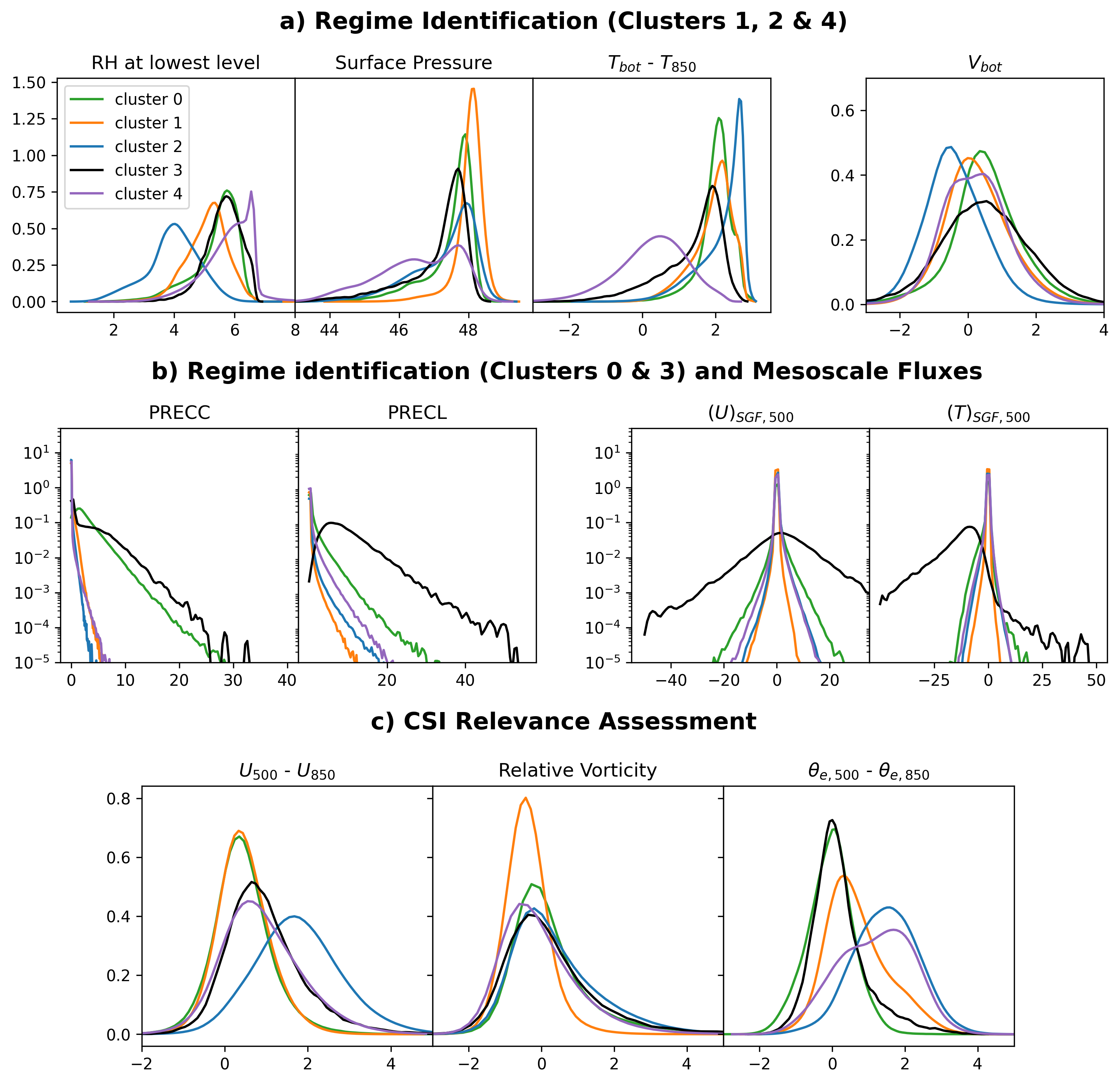}
    \caption{Kernel density estimations for the cluster distribution of a few relevant features used in the K-means++ clustering algorithm  as well as additional important variables which are not used for the clustering. a) includes features and variables that especially help to identify clusters 1, 2 and 4, b) contains the distributions of precipitation and the mesoscale fluxes of zonal wind and heat, which help identify clusters 0 and 3, and c) contains variables that help assess the relevance of CSI in the different clusters.
    The densities are normalized separately for each cluster, and each variable is standardized (but not centered) independently. Thus, the units in the x-axis are in standard deviation of the variables along the sample dimension. Note the logarithmic scales for precipitation and the subgrid-scale fluxes. 
    }
    \label{fig:distributions_kmeans}
\end{figure}

The distributions in Figures \ref{fig:distributions_kmeans}a and \ref{fig:distributions_kmeans}b are used to identify regimes according to their main characteristics, and Figure \ref{fig:intro_plot}a shows the spatial distribution of clusters within an example winter extratropical cyclone, illustrating the interpretation of the clustering as described below. For a clearer overview, a list with a description of the clusters is presented in Table \ref{tab: cluster characterization} in the supporting information.

We identify clusters 1, 2, and 4 using the marginal distributions depicted in Figure \ref{fig:distributions_kmeans}a.  Cluster 1 contains 45\%  of the samples and is the largest. It is characterized by notably high surface pressure values and contains fewer high mesoscale fluxes than the other clusters. This cluster indicates a relatively stable atmosphere with reduced mesoscale disturbances and is the least relevant in terms of mesoscale activity; it is therefore labeled as \textit{Stable}. Cluster 2 shows characteristics of high values of $T_{\text{bot}}-T_{850}$ and low moisture values compared to the other clusters, indicating a dry regime with cold air over a warm near-surface layer.  We name this cluster \textit{CAO} because its spatial distribution suggest that it contains cold air outbreaks and the cold sector of cyclones (Figure \ref{fig:intro_plot}a). This is also supported by the distribution of meridional velocity at the bottom level $V_{\text{bot}}$, where cluster 2 contains the most northerly wind values. Cluster 4, on the other hand, is a moist regime, with the lowest pressure values and warmest temperatures at 850 hPa. This cluster is therefore identified as \textit{Warm moist air masses}, including parts of the warm sector of cyclones. Consistently, this cluster also has more precipitation from resolved motions (PRECL) than clusters 1 and 2 (Figure \ref{fig:distributions_kmeans}b).

Clusters 0 and 3 are identified with the help of the distributions in Figure \ref{fig:distributions_kmeans}b. Cluster 3 is the smallest cluster, comprising only 0.3\% of the samples, and has the largest subgrid fluxes as well as a large amount of CAPE and strong convergence at the lowest level (Figure \ref{fig:distributions_kmeans_all} in Supporting Information). Cluster 0 is similar to cluster 3, but it contains more samples and subgrid-scale fluxes that are not as extreme in magnitude. The resolved and parameterized precipitation distributions clarify what mainly distinguishes these two clusters. Cluster 3 contains large values of PRECL, which is the precipitation coming from resolved motions, as for example in fronts or resolved convection. It also contains the largest values of parameterized convective precipitation (PRECC) and is therefore related to extreme weather and heavy precipitation. Cluster 0 on the other hand contains similarly large PRECC but much smaller PRECL. We therefore identify cluster 0 as the regime of \textit{Precipitating Convection} and cluster 3 as the regime of \textit{Organized heavy precipitation}.

The distributions in Figure \ref{fig:distributions_kmeans}c help to assess which of the clusters capture slantwise convection.
Clusters 2 and 4 have a combination of high vertical shear of zonal wind $U_{500}-U_{850}$ and positive values of the equivalent potential temperature difference
$\theta_{e,500}-\theta_{e,850}$. This combination of characteristics is conducive to CSI and slantwise convection, but not as much to upright convection. 
This can be understood based on the literature which identifies CSI through negative moist potential vorticity (MPV):
\begin{equation}
    \text{MPV} = \frac{1}{\rho}(\boldsymbol{\eta}\cdot \nabla\theta_e) = \frac{1}{\rho}(f + \zeta_{\theta_e})\left.\frac{\partial \theta_e}{\partial z}\right|_{\theta_e}
    \label{eq:MPV}
\end{equation}
where $\theta_e$ is the equivalent potential temperature, $\boldsymbol{\eta}$ the absolute vorticity vector and $\rho$ the air density. The second form in equation \ref{eq:MPV} is expressed in isentropic coordinates involving the Coriolis parameter $f$ and the relative vorticity on isentropic surfaces $\zeta_{\theta_e}$. 
In the Northern Hemisphere, potential vorticity (PV) is usually positive. MPV can become negative by either switching sign in the vertical derivative of equivalent potential temperature or by strong negative values of relative vorticity on isentropic surfaces. The former would lead to conditional 
instability, whereas symmetric instability is rather caused by strong negative values of isentropic relative vorticity. In a baroclinic zone, where isentropes slope upwards and polewards, large positive values of $U_{500}-U_{850}$ indicate a shear along isentropes that can flip the sign of $\zeta_{\theta_e}$ and thereby increase the probability of negative MPV.
The connection to slantwise convection is supported by the statistics of the zonal momentum flux distribution for cluster 2 (Table \ref{tab:cluster-stats-arbitrary} in Supporting information), which has a positive skewness and large proportions (around 60\%) of positive values. These statistics indicate that cluster 2 contains a large amount of downward zonal momentum fluxes in the mid-troposphere (downward fluxes are positive because $\omega$ in pressure coordinates is positive downward), whereas this is somewhat less true for cluster 4. CSI requires moisture, and the reason cluster 2 is relevant to slantwise convection despite its low near-surface relative humidity (Figure \ref{fig:distributions_kmeans}a) is that when the cold air passes over the warm, moist Atlantic Ocean, it can gain moisture through surface fluxes. 
The regime in cluster 3 is also expected to be relevant for CSI, since regions of high slantwise CAPE (SCAPE) encompass regions of high CAPE, but the distribution in $\theta_{e,500}-\theta_{e,850}$ suggests that this cluster is leading primarily to upright and less to slantwise convection and is associated with cold fronts and point storms.

\subsection{Performance of the ANN}
\label{subsec:performance results}

\begin{figure}[h]
    \centering
    \includegraphics[width=0.8\linewidth]{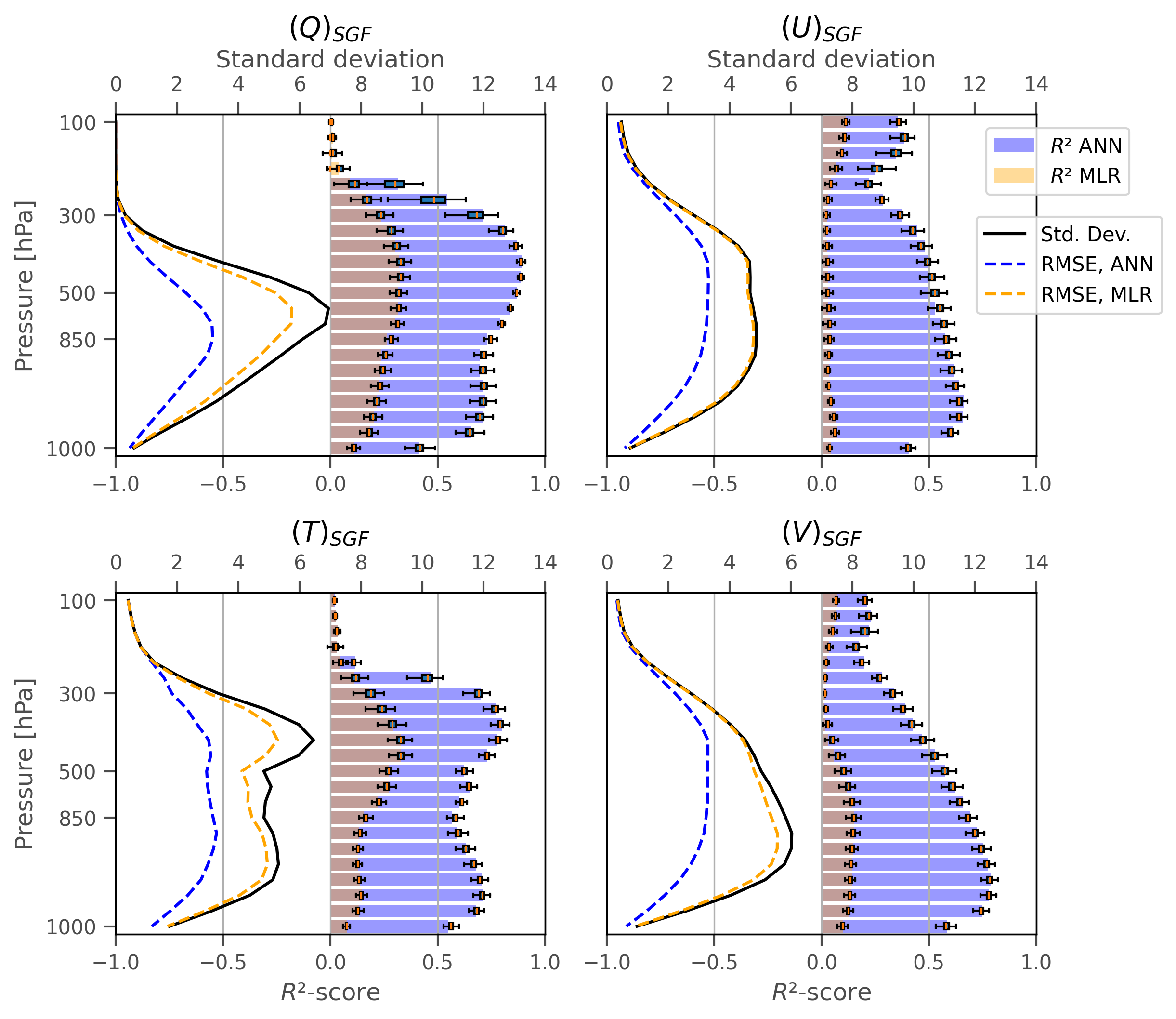}
    \caption{The main ANN model is skillful for profile-to-profile mappings.
    The bars represent the $R^2$ scores for every atmospheric level, where blue is the skill of the ANN and orange is the skill of the multiple linear regression (MLR) baseline. Additionally, for each level, the standard deviation of the test data is displayed in black, as well as the root mean square error (RMSE) of the ANN in blue and the RMSE of the MLR in orange. The units for these metrics are arbitrary due to the normalization. Uncertainty estimation is presented as a boxplot indicating the median, the interquartile range, and the whiskers of the distribution obtained by bootstrapping. The average pressure of the hybrid-coordinate vertical levels is shown on the y-axis.}
    \label{fig:performance}
\end{figure}

In this section, we evaluate the performance of the ANN trained with all input features for the profile-to-profile mapping. The predictive skill is estimated using the coefficient of determination $R^2$ evaluated on the test set and indicates the extent to which the ANN can explain the variance of the targets. The resulting $R^2$ scores for the four subgrid-scale flux variables are presented in Figure \ref{fig:performance}. Overall, the highest $R^2$ values are observed for the moisture flux across most levels. For the uppermost levels, the standard deviation of the moisture fluxes approaches zero. This results in negative $R^2$ scores due to the denominator in the definition of the metric:
\begin{equation}
    R^{2} = 1 - \frac{\sum_i (y_{\text{true,}i} - y_{\text{pred,}i})²}{\sum_i (y_{\text{true,}i} - \bar{y}_{\text{true}})²} = 1 - \frac{\text{MSE}}{\text{Data Variance}}.
    \label{eq:R^2-definition}
\end{equation}
Here $y_{\text{pred,}i}$ are the predictions of the ANN for each sample $i$ (consisting of different time steps, latitudes and/or longitudes, but can contain also different levels, for example when computing a global $R^2$ score or displaying the score on a map)  while $y_{\text{true,}i}$ are the true targets with mean $\overline{y}_{\text{true}}$ across sample dimension. The fraction in equation \ref{eq:R^2-definition} can be interpreted as the ratio of the mean square error (MSE) to the variance of the true targets. If both of these quantities in the fraction are close to zero, the metric evaluates noise, and negative values become possible. To counteract this issue, the moisture flux output of the ANN in the five uppermost levels has been explicitly set to zero, replacing the prediction with surrogate statistics and resulting in zero $R^2$ scores at these levels. For similar reasons, the score is lower at the lowest level for each of the flux variables.

Among the subgrid-scale fluxes, the momentum fluxes are the most challenging ones to predict. However, they demonstrate better predictability at the upper levels than moisture and heat fluxes. The best predicted levels vary between the fluxes: While the momentum fluxes are best captured in the lower part of the troposphere, the moisture flux is better predicted in the mid-troposphere. 

The regional variation in $R^2$ scores is provided in Figures \ref{fig: r2 altitude latitude} and \ref{fig: r2 map} in the Supporting Information and reveals a range of values spanning from negative values at some spots to near-perfect predictions with $R^2$ scores approaching 1. Negative values are rare but appear in the free troposphere at latitudes between 30$^{\circ}$N and 35$^{\circ}$N in the prediction of heat fluxes and some isolated regional spots for the momentum fluxes. Notably, the vertical variation of the $R^2$ score at each latitude is stronger for the heat and moisture fluxes than for the momentum fluxes, with a great drop in performance for levels above roughly 200 hPa. In addition, the ANN performs better over the ocean than over land (Figure \ref{fig: r2 map}). 

The benefit of using nonlinear models is highlighted by comparing the performance of the ANN with a multiple linear regression (MLR) baseline (Figure \ref{fig:performance}). The ANN considerably outperforms the MLR, in particular for the momentum fluxes. 
The root mean square error (RMSE) of the MLR follows the shape of the standard deviation of the data more closely and performs only slightly better than the surrogate statistics.

\begin{figure}[h!]
    \centering
    \includegraphics[width=1\linewidth]{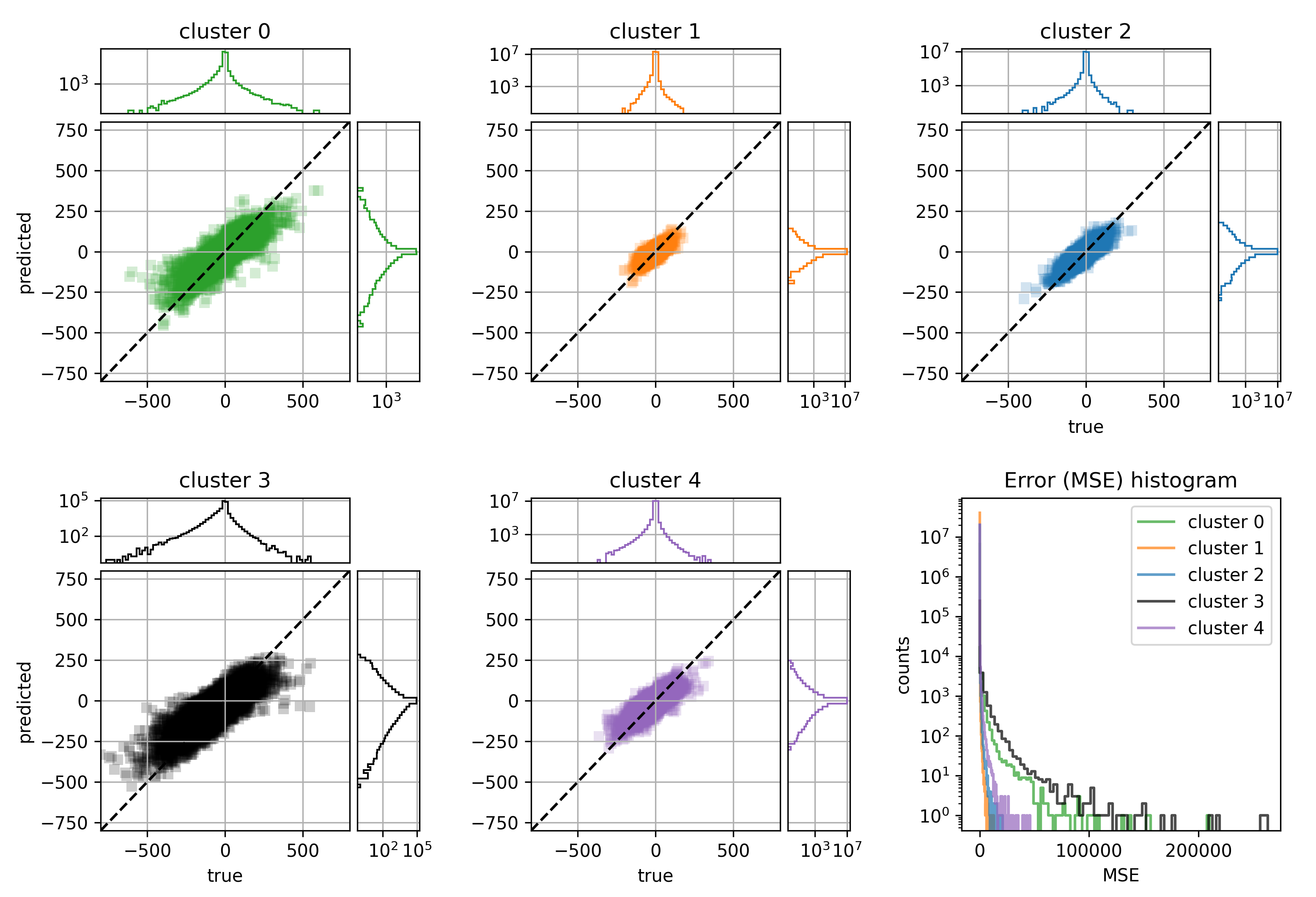}
    \caption{Predicted vs. true scatter plots for Clusters 0-4. Each data point represents an output of a sample (any one of the subgrid fluxes from any level).
    The dashed diagonal line indicates the ideal case where predictions would perfectly match the true values. On the top marginal of the scatter plot, the histogram of the true values is displayed, while on the right marginal, the histogram of the predicted values is displayed. Both histograms show the counts in logarithmic scale.
    In the bottom right panel, the histograms of the mean square error for each cluster are displayed with counts in logarithmic scale.}
    \label{fig: scatter predictions true linear}
\end{figure}

The model selection procedure with Optuna revealed that one of the main hyperparameters influencing the skill of the model is the loss function and the best performing models use the balanced L1 loss function. This loss reduces sensitivity to outliers by fitting the bulk of the distribution over extremes. In contrast, an MSE loss would target the conditional mean given the inputs and weights large residuals quadratically, which can overemphasize a small number of extreme samples when the residual distribution is heavy-tailed. As a result, the balanced-loss ANN shows larger errors in high-variance regimes than in low-variance regimes, even for intermediate values. In the predicted vs. true scatter plots (Figure \ref{fig: scatter predictions true linear}), the predicted values for the two precipitating clusters (0 and 3) are more widely dispersed around the true values. This suggests that, in these regimes, the ANN recognizes the need to produce larger values but struggles to consistently determine the exact conditions under which they should occur. In all regimes, the slope of the predicted values is smaller than one, implying that the ANN tends to underestimate large absolute values of the subgrid fluxes. 
Cluster 3 has by far the largest errors with 54\% of the predictions exceeding the 95th percentile of the MSE across all of the test sets' samples and outputs ($\text{MSE}_{95\%} = 7.18$). As a comparison, the samples of clusters 0, 1, 2 and 4 exceed the $\text{MSE}_{95\%}$ threshold only 12\%, 2\%, 6\% and 7\% of the time, respectively.

We examine the frequency with which the predicted sign of an output exceeding the $\text{MSE}_{95\%}$ threshold matches the true sign.
This is a crucial metric, as the sign conveys information about the direction of the underlying processes, such as up- and downdrafts in convection. Examples of such misclassifications can be observed in the extra tropical cyclone in Figure~\ref{fig:intro_plot} by comparing the true subgrid-scale heat fluxes and the predicted ones in panels e and f. One such example is the downward flux located roughly at 56°W and 36°N along the frontal feature, which is predicted as a upward flux by the ANN. 
In the predicted vs. true plots (Figure \ref{fig: scatter predictions true linear}), the ANN correctly predicts the sign for values in the first and third quadrants, while it incorrectly predicts the sign for values in the second and fourth quadrants. Among the different clusters, clusters 0 and 3 exhibit the highest percentage of correctly predicted signs in large MSE samples, with 73$\%$ and 78$\%$, respectively. In contrast, clusters 1, 2 and 4, which are characterized by lower variance in the flux distributions, have slightly lower percentages of correctly predicted signs in large MSE samples, at 67$\%$, 66$\%$, and 65$\%$, respectively. In these clusters, the severity of falsely predicted signs is lower as the magnitudes of the flux values are smaller. This analysis suggests that extreme deviations from the true values in the wrong direction tend to be relatively rare.

\subsection{Which Meteorological Variables Contribute Most to Model Skill?}
\label{subsec:results feature importance p1}

Ablation experiments are conducted based on removing categories of input features, following a hierarchy motivated by physical interpretation, which is summarized in Table \ref{tab: Hierarchy categories}. Category 0 contains the most fundamental dry variables. They are expected to explain many of the dynamics on their own as they are the building blocks for many other atmospheric variables. Category 1 includes horizontal temperature gradients which, if thermal wind applies, can contain redundant information when used together with category 0. Other horizontal neighboring information is included in category 2 and 3, containing horizontal derivatives of momentum, where category 2 contains the horizontal divergence terms and category 3 contains the vorticity terms. The effect of moist dynamics is examined by the inclusion of category 4. Besides $Q$, category 4 also contains CAPE, which uses moisture information in its computation.  
The last category includes the vertical hybrid pressure-sigma velocity $\omega$, which is treated separately as there are concerns that the coarse-grained vertical velocity might be influenced by smaller scale dynamics such that it might not be equivalent to vertical velocities from a low-resolution simulation. This category is used to compare ANN performance with and without it. 

\begin{table}[h]
    \centering
    \caption{\small Hierarchy of the categories involved in the construction of the ablation experiments.}
    \begin{tabular}{ p{2cm}|p{2cm}|p{8cm} } 
    \hline
     \bf Hierarchy & \bf Variables & \bf Physical Characterization\\ 
     \hline
     Category 0 & $U,\ V,\ T,\ P_S$ & Fundamental dry variables\\ 
     Category 1 & $\partial_x T$, $\partial_y T$  & Horizontal temperature gradients \\ 
     Category 2  & $\partial_x U$, $\partial_y V$ & Terms for computing the divergence\\ 
     Category 3  & $\partial_x V$, $\partial_y U$ & Terms for computing the relative vorticity \\ 
     Category 4 & $Q$, CAPE & Variables involving moisture \\ 
     Category 5 & $\omega$  & Vertical velocity\\ 
     \hline
    \end{tabular}
    \label{tab: Hierarchy categories}
\end{table}

\begin{figure}[h]
    \centering
    \includegraphics[width=1\linewidth]{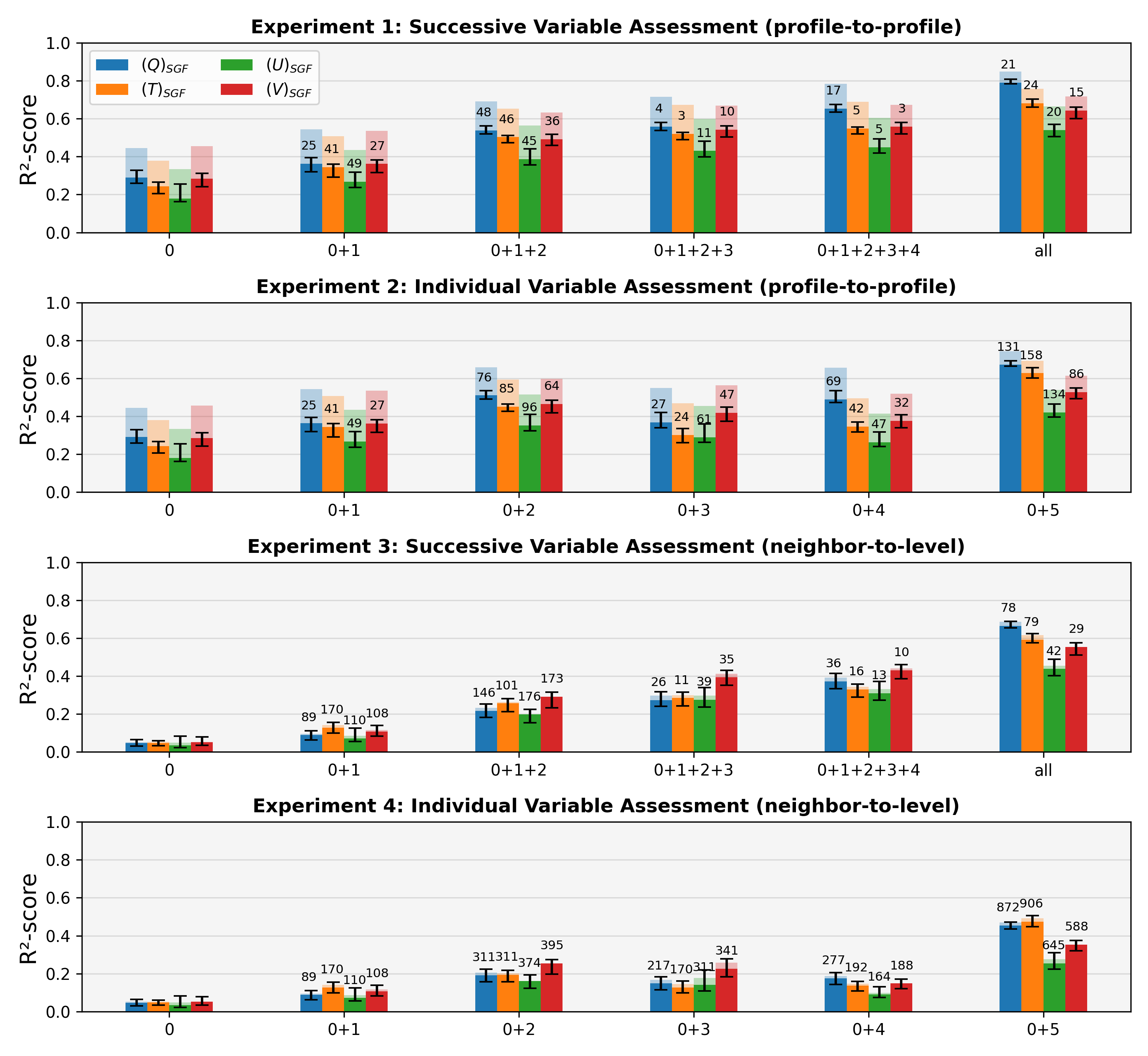}
    \caption{Ablation experiments for the assessment of variable feature importance. Each set of four bars corresponds to the performance of different models trained with varying subsets of features. The performance is quantified by the $R^2$ score, averaged over the vertical profile for each of the four subgrid-scale fluxes. The bars in lighter colors are the $R^2$ values for the training and visualize the generalizability of the models.
    Experiments 1 to 4 focus on assessing variable importance: Experiments 1 and 2 are conducted using a profile-to-profile mapping, while Experiments 3 and 4 use a neighbor-to-level mapping. The x-axis shows which feature categories are included in each ANN model. Experiments 1 and 3 use a successive assessment of category importance according to the hierarchy outlined in Table \ref{tab: Hierarchy categories}. The numbers over the bars in experiments 1 and 3 represent the performance improvement in percent for a model containing categories 0 to $i$ (where $i\in \llbracket 1,5\rrbracket$) with respect to the performance of a model containing categories 0 to $i-1$. Experiments 2 and 4 assess the importance of categories 1 to 4 by adding them individually to category 0; the numbers on top of the bar for these experiments are with respect to the model trained only with category 0.}
    \label{fig:ablation_experiments_0to4}
\end{figure}

The first ablation experiment, shown in the top panel of Figure \ref{fig:ablation_experiments_0to4}, quantifies the skill gained in a profile-to-profile mapping when variable categories are added successively according to the hierarchy outlined in Table \ref{tab: Hierarchy categories}. For each new model in the experiment, the improvement is expressed as a percentage and is obtained relative to the performance of the model without the newly added category. For example, if a model includes categories 0 to 3 as inputs, the values displayed on top of the bars represent the improvement compared to a model using only categories 0 to 2. 
Adding the categories successively according to the aforementioned hierarchy will diminish the added value of categories that are added later, given that the features may contain correlations with each other, and the model will have already learned most of what can be learned from this shared feature variance. Therefore, we conduct a second ablation experiment where categories 1 to 5 are individually added to category 0. The second panel of Figure \ref{fig:ablation_experiments_0to4} expresses the improvement from adding features in each category relative to the baseline model using only the variables in category 0 as features. We need to keep in mind that this experiment is still not entirely fair to all the categories as they do not consist of equal number of features (e.g., categories 4 and 5 contain a smaller number of features than other categories).

Out of categories 1 to 4, the divergence terms in category 2 are particularly important for predicting all fluxes. Moisture and heat fluxes also benefit greatly from the inclusion of the moisture variables in category 4. The contribution of moisture is diminished in experiment 1, where it is included after the dry variables. The only flux variable that experiences a significant improvement with the inclusion of category 4 in experiment 1 is the moisture flux. In the individual variable assessment (experiment 2), both momentum fluxes profit the most from the inclusion of the divergence and vorticity terms.

The improvements seen with the inclusion of category 1 suggest that centered horizontal differences in temperature are not redundant with the thermal wind captured by the zonal and meridional wind profiles in category 0. Adding category 1 has a notably high impact on the performance of the zonal momentum flux, with an improvement of 49\%.  These improvements are likely due to the fact that thermal wind breaks down on the small scales that we are considering. Another plausible explanation is that in practice the landscape of the loss function in a neural network does not allow the global minimum to be reached and therefore adding information that is redundant in theory can still improve the performance of the ANN.

\begin{figure}[h]
    \centering
    \includegraphics[width=0.99\linewidth]{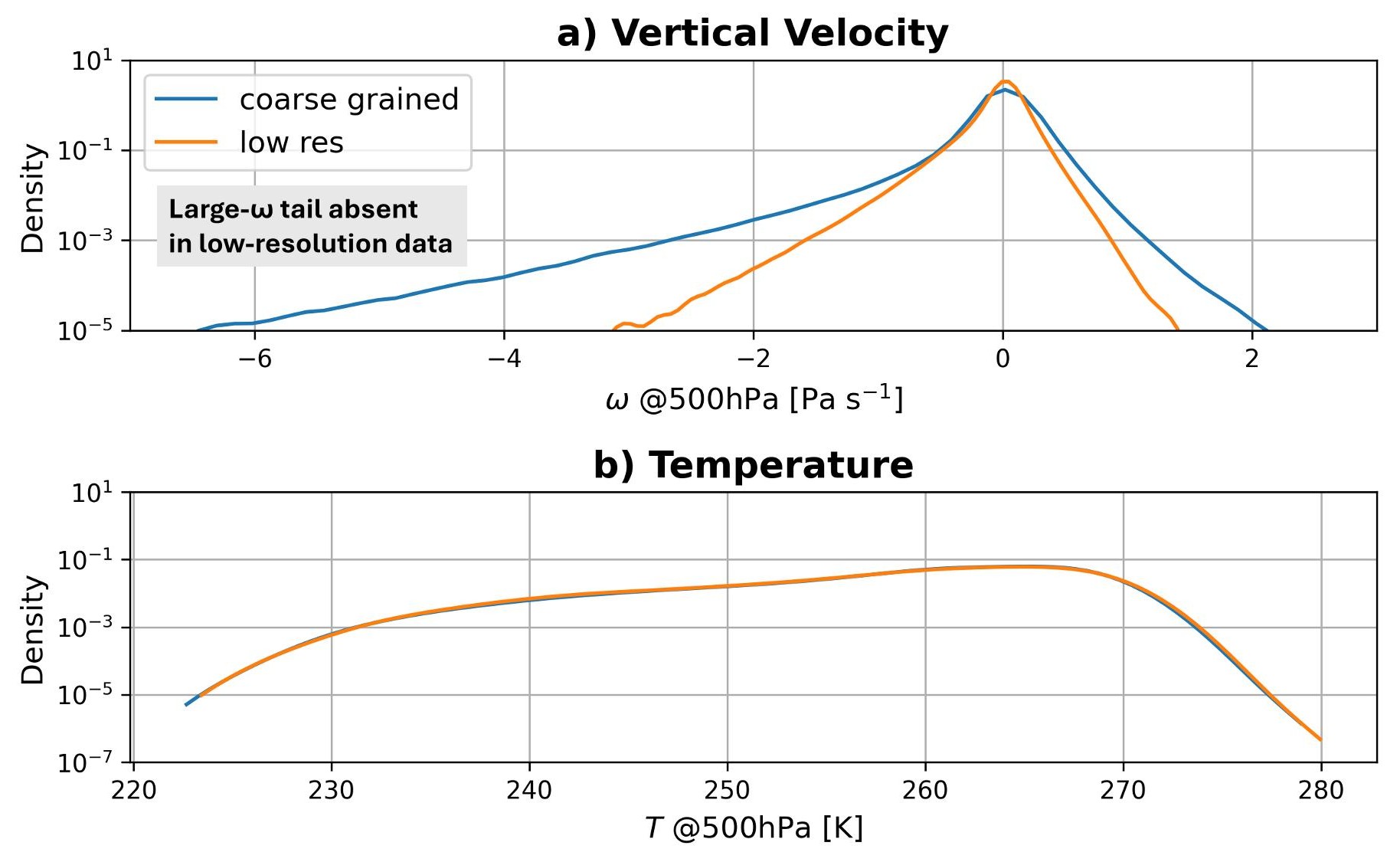}
    \caption{A comparison of the kernel density estimation of the distributions of coarse-grained high-resolution $\omega$ and $\omega$ from a comparable low-resolution simulation. The bottom panel shows the same comparison for the temperature. The logarithmic scale was chosen to highlight the different behavior at the tails between the two variables.}
    \label{fig:comparison_omega}
\end{figure}

The prediction of each subgrid flux variable is significantly improved by the inclusion of the vertical velocity as a feature. In experiment 1, adding vertical velocity in the final step improves ANN performance by about 20\% for all flux variables, highlighting that it provides additional predictive information even when divergence is included. Additionally, experiment 2 shows that the individual contribution of $\omega$ is significantly higher than that of the divergence terms. Since category 5 has half as many features as category 2 there is a much higher skill improvement per feature. 
This is the first hint that the vertical velocity adds much more value in a setting where the features are coarse-grained from higher resolutions than it would when vertical velocities are taken from a coarse-resolution model, i.e., that vertical velocity coarse grained from a high-resolution simulation may contain more information than vertical velocity in a low-resolution simulation. 

In Figure \ref{fig:comparison_omega}, we compare the distributions of the coarse-grained variables $\omega$ and $T$ with those obtained from low-resolution simulations (110-km grid spacing) that are otherwise comparable to the high-resolution simulations, differing only in horizontal resolution (\cite{wills_resolving_2024}). Figure \ref{fig:comparison_omega}a validates that coarse-grained $\omega$ is significantly different from low-resolution $\omega$ because the distributions differ substantially. In particular, there are vertical velocities of larger magnitude when $\omega$ is coarse-grained, with the maximum upward velocities more than twice as large. This is not the case for other features, such as the temperature in Figure \ref{fig:comparison_omega}b, where the distributions are almost identical for the low-resolution variable and the coarse-grained variable. This indicates that coarse-grained $\omega$ likely encodes information about when large subgrid-scale vertical velocities and fluxes are occurring. 

To compare the importance of the various feature categories in a local mapping, where the ANN only receives information about neighboring levels, we repeat the ablation experiments that had been conducted for the profile-to-profile mapping for the neighbor-to-level mapping (Experiments 3 and 4 in Figure \ref{fig:ablation_experiments_0to4}). Category 0 alone performs poorly in a local mapping, with $R^2$ scores below 0.1 for all output variables and is only slightly better than using surrogate statistics. Adding any of the other categories to category 0 increases the skill by over 100\% in most cases (Experiment 4), suggesting that in the case of a local mapping, the features in categories 1, 2, 3, and 4 are more powerful predictors than the fundamental dry variables in category 0. 

Notably, the vertical velocity captures nearly the entire skill of a neighbor-to-level mapping when added individually to category 0. In this case, the improvement factors are huge, with between 588\% ($(V)_{\text{SGF}}$) and 906\% ($(T)_{\text{SGF}}$) improvements. Adding the vertical velocity to categories 0-4 in experiment 3 improves the performance by up to 79\%. This substantial impact of $\omega$ on model performance further validates the suspicion that it contains too much information when coarse-grained from a high-resolution model output and might be less informative in an online setting within a coarse-resolution ESM.

The substantial drop in performance when using a local mapping with a reduced number of features poses a challenge for equation discovery through ML techniques, as a reduced number of features is crucial in symbolic regression.
Since the divergence terms are relatively good predictors in cases where there is a reduced number of features, we do an additional experiment comparing the performance of a neighbor-to-level mapping for a model that uses category 0 and 2 and a model that uses category 0 and one feature which combines the two components of divergence into a single term. We find an average drop of 25\% in performance across the four target fluxes when the two components are combined into a single term. This suggests that the ML model learns better from the individual terms than from the combined divergence term. 

\begin{figure}[h]
\includegraphics[width=1.0\linewidth]{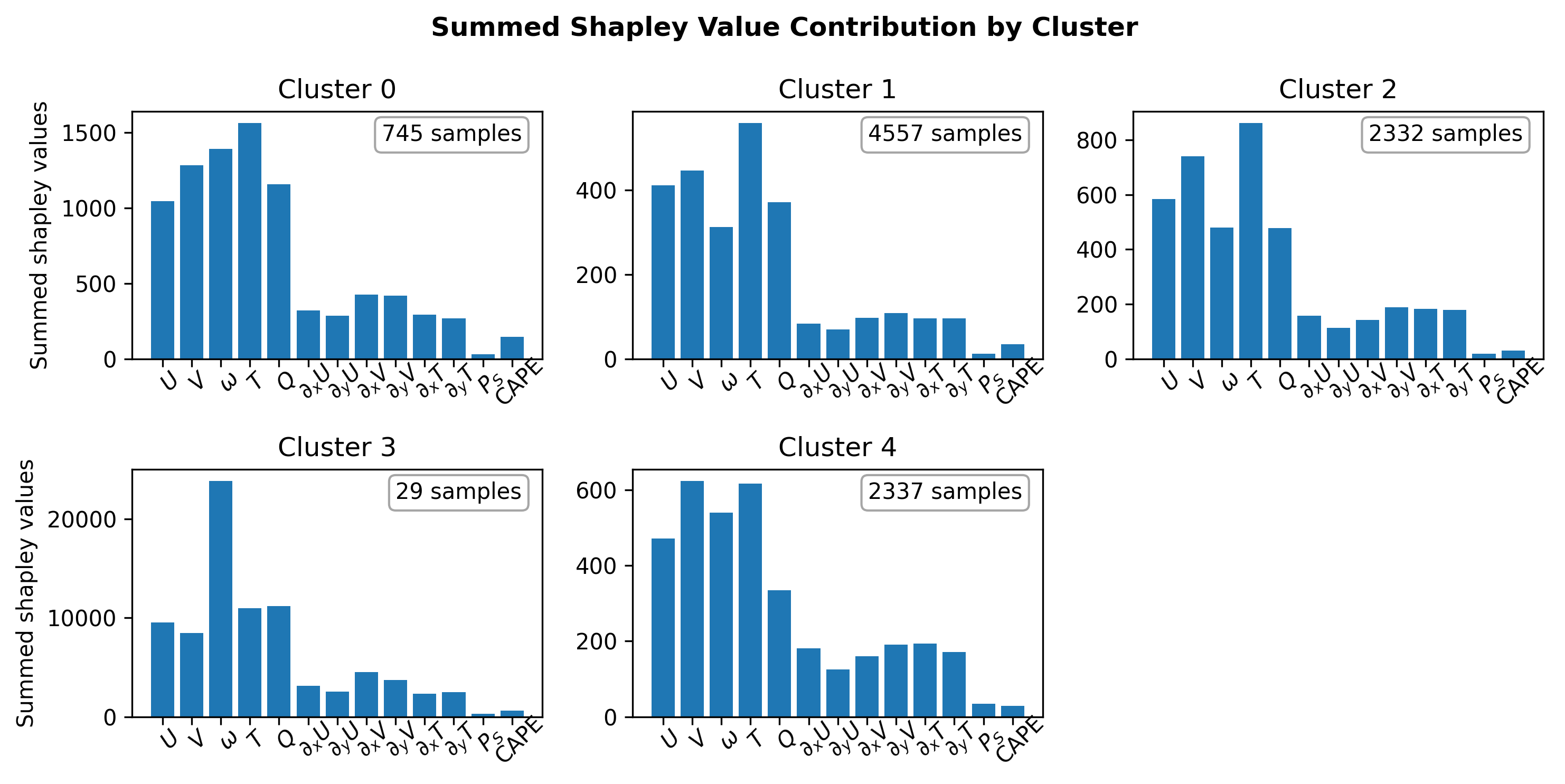}
    \caption{Shapley analysis of the net effect of the meteorological variables on all output fluxes by cluster. The mean absolute Shapley values are summed over all levels of the variable and over all levels of all four output fluxes. To obtain the mean absolute Shapley values, the summation is performed across the sample dimension. Ten batches of 1000 test samples are separated according to their cluster label. The results presented are the mean across the samples in all 10 batches. The number of the samples in each cluster across the 10 batches is indicated in the text box.}
    \label{fig:shapley sum cluster}
\end{figure}

The summed absolute Shapley values for the total impact of a variable on the output variables (Figure \ref{fig:shapley sum cluster}) provide insight into which variables the main ANN (the profile-to-profile ANN mapping with all features included) has learned from. While the ablation experiment hinted at a high importance of the vertical velocity and horizontal derivatives, the Shapley values tend to be relatively small for the vertical velocity and extremely small for the horizontal derivatives. The summed absolute Shapley values for the different clusters show that vertical velocity is most important in the regime of \textit{Organized heavy precipitation} in cluster 3. This cluster has the largest subgrid-scale fluxes, and therefore, the largest Shapley values are assigned to this cluster. However, it also contains the smallest number of samples. As a result, the effect of vertical velocity on the model performance is large, but the average Shapley value across all clusters is small. This explanation cannot be made for the discrepancy in the importance of the divergence terms between the Shapley analysis (Figure \ref{fig:shapley sum cluster}) and the ablation experiments (Figure \ref{fig:ablation_experiments_0to4}). 
Here, we speculate that in the presence of the moisture and vertical velocity features, the ANN assigns large weights to these features, resulting in lower weights and hence lower Shapley values for the horizontal derivative terms because they might be highly correlated with each other.

\subsection{Vertical Localization Relationship}
\label{subsec:results feature importance p2}

\begin{figure}[h]
    \centering
    \includegraphics[width=1\linewidth]{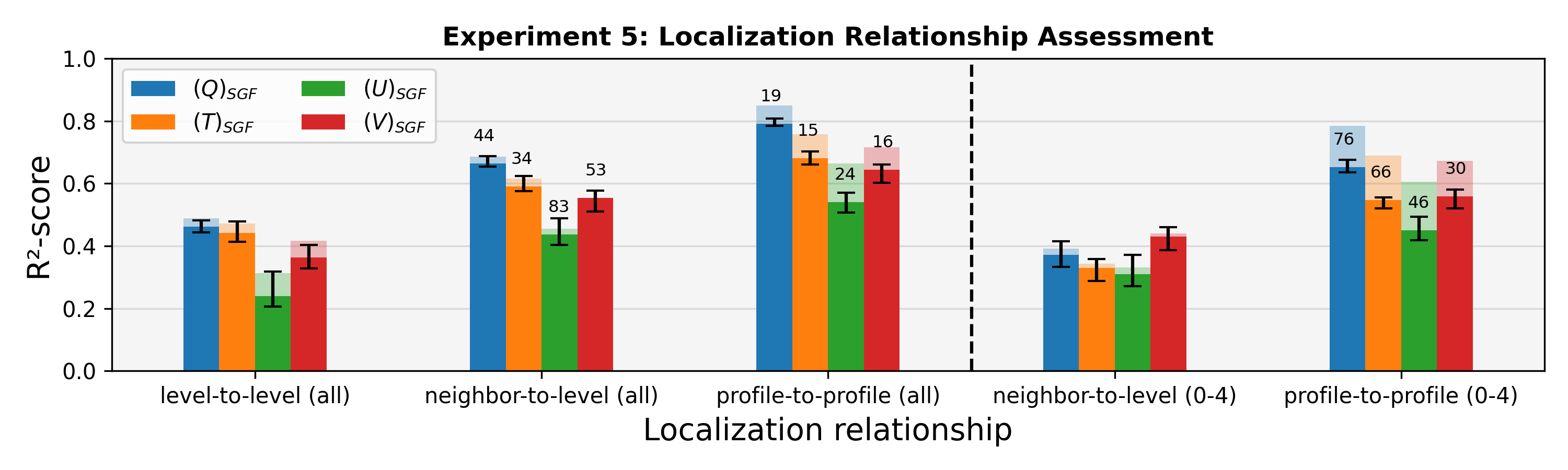}
    \caption{Ablation experiment 5 focuses on the assessment of localization relationship importance. Left of the vertical dashed line is a comparison for three localization relationships: level-to-level, neighbor-to-level, and profile-to-profile, when all features inclusive $\omega$ are included. The improvement percentages of the neighbor-to-level model are with respect to the level-to-level performance, while the profile-to-profile improvement is measured with respect to the neighbor-to-level performance. Right of the vertical dashed line is a comparison of the neighbor-to-level and profile-to-profile performance without the inclusion of the vertical wind as an input feature, and the improvement percentage of the profile-to-profile mapping is indicated.}
    \label{fig:ablation_experiments_5}
\end{figure}

Experiment 5 in Figure \ref{fig:ablation_experiments_5} compares the performance of neural networks with different vertical localization relationships. When all features are included, the gain in performance in a neighbor-to-level mapping relative to a level-to-level mapping is more pronounced than the gain in a profile-to-profile relative to a neighbor-to-level mapping. When excluding the vertical velocity as a feature, the performance gain in a profile-to-profile mapping becomes more important, where especially the moisture and heat flux predictions profit from a non-local mapping.

\begin{figure}[h]
    \centering
    \includegraphics[width=0.65\linewidth]{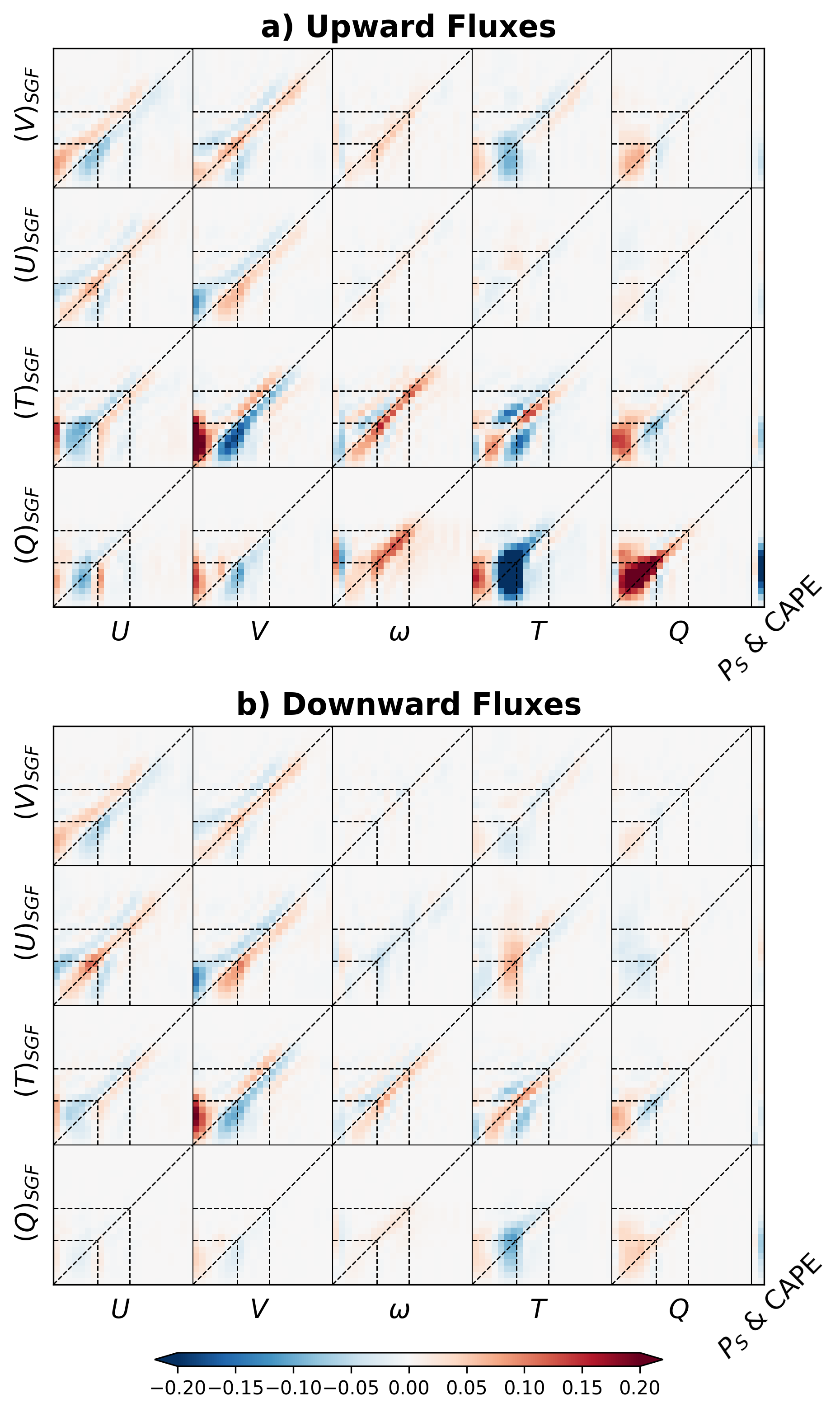}
    \caption{Shapley response matrices illustrating the level-wise importance of features for the profile-to-profile mapping for upward fluxes. These nonlinear response matrices display the sign information of the Shapley values. The x-axis represents a selection of features ($U,\ V,\ \omega,\ T,\ Q,\ P_S,$ CAPE), with lower levels positioned further to the left, while the y-axis represents the targets, with lower levels positioned further below. The 850 hPa and 500 hPa lines are indicated as horizontal and vertical lines. Results are averaged across ten batches; for each batch, 1000 subsamples are drawn at random. On the top, only subsamples exhibiting upward fluxes contribute to the Shapley-value mean, while on the bottom plot, only downward fluxes are included. The Shapley values have the same units as the outputs of the ML model, which in our case are arbitrary due to their normalization.}
    \label{Figure:shapley}
\end{figure}

The Shapley response matrices for the profile-to-profile mapping in Figure \ref{Figure:shapley} provide further intuition about the nature of the localization relationship between the outputs and inputs and reveal what information may be missing in cases where there is a significant performance gap between a vertically local vs. non-local mapping. 
Across all variables, the ANN appears to learn relationships in the boundary layer, as can be seen from large values in the bottom left of each panel. On the other hand, there are vanishing Shapley values in the uppermost levels. This is likely a byproduct of the fluxes approaching zero at upper levels and therefore not being influenced when features at upper levels are perturbed. In general, small Shapley values can also indicate a stochastic contribution if there exists a high uncertainty in the relationship between the features and outputs.
The largest Shapley values are found primarily along and close to the diagonal of the matrix, which represents the same vertical levels in both the input and output variables. This suggests that the localization relationship is largely local / quasi-local.

Nonetheless, there are some significant non-local effects of the features on the prediction. Features at levels near the surface are found to be important for predicting subgrid-scale fluxes throughout the boundary layer. The importance of non-local effects within the boundary layer is particularly visible in the $Q-(Q)_{\text{SGF}}$ matrix, which shows how specific humidity influences the prediction of sub-grid moisture fluxes. Changes in moisture within the boundary layer have a strong non-local effect on moisture flux, where a change in moisture in one level modifies the moisture flux near and above that level. CAPE also has a strong non-local effect on the moisture fluxes. 
Temperature has even a stronger vertically non-local effect, which is expected since the vertical profile of temperature characterizes the atmospheric stability. 
Therefore, temperature information from one level can influence fluxes across a broad vertical range. 
Meridional momentum close to the surface also has a strong non-local effect, particularly on the heat fluxes, but also to a significant extent on the moisture and zonal momentum fluxes. Meridional wind at higher levels mainly influences fluxes at levels near the diagonal, i.e., quasi-locally.
As expected from the ablation experiments, vertical velocity shows a more local relationship to the target fluxes and is mainly important for the prediction of heat and moisture fluxes.
Certain mesoscale fluxes in the free troposphere do not depend on momentum at the same level but instead on neighboring levels around the diagonal. This is largely true for the $U-(V)_{\text{SGF}}$,  $V-(U)_{\text{SGF}}$,  $U-(T)_{\text{SGF}}$, and  $V-(T)_{\text{SGF}}$ matrices. Such structures could explain why performing a neighbor-to-level mapping might have led to a strong improvement in the skill compared to a level-to-level mapping. 

The response matrices for the horizontal derivative features (Figure \ref{Figure: SHAP sign matrices}) show significantly smaller Shapley values but several robust non-local effects. For example, a clear neighbor-to-level response of the divergence terms on the mesoscale heat flux is apparent. The derivatives of the meridional momentum in the boundary layer (e.g., associated with abrupt wind direction changes in fronts) have stronger non-local effects on mesoscale fluxes than other horizontal derivative features. At higher levels, the horizontal derivatives of temperature in particular demonstrate a non-local connection to the mesoscale fluxes.


\subsection{Physical Insights from Feature Importance Analysis}
\label{subsec:results feature importance p3}

The Shapley responses can also give physical insights into the dynamics governing mesoscale fluxes in the front-resolving simulations, although some are more intuitive than others. For example, there are straightforward relationships between low surface pressure and a higher likelihood of upward fluxes and between increased CAPE and a higher likelihood of upward fluxes. Keeping in mind that the behavior of an ANN does not always reflect the underlying physical mechanism and that it sometimes misclassifies the sign of the fluxes, we try to make sense of some of the other notable and explainable characteristics in the Shapley response matrices (Figures \ref{Figure:shapley} and \ref{Figure: SHAP sign matrices}). The Shapley matrices are computed separately for cases where the true fluxes are directed upward and downwards. Since upward fluxes are negative, negative values in the Shapley matrices indicate conditions conducive to upward fluxes. For example, the large blue spot in the $T-(Q)_{SGF}$ matrix indicates warm temperatures at the top of the boundary layer can help to predict upward sub-grid moisture fluxes throughout the boundary layer. 

The Shapley $Q-(Q)_{SGF}$ matrix has larger values for upward than downward fluxes because moisture fluxes have a strong negative skewness and high percentage of negative values in the moisture flux distribution (Statistics in Supporting Information \ref{tab:stat by cluster}).
This is consistent with vertical motions being moist when directed upwards and dry when directed downward.
The Shapley $Q-(Q)_{SGF}$ matrix shows a reduction of upward moisture fluxes when there is increased moisture in the atmosphere. This can be explained by a larger saturation deficit reducing condensation such that there can be continued upward moisture transport at the level of the moisture deficit and the levels above it. 

A similar explanation as for the $Q-(Q)_{SGF}$ matrix applies to the strong negative response in the $T-(Q)_{SGF}$ plot: saturation specific humidity increases with temperature such that the atmosphere can hold more moisture before condensation occurs. In cases where there is updraft, this would mean increased upward moisture transport for warmer temperatures, because the moisture is transported further upwards before condensing.
Warming in the upper boundary layer usually means that there is broader warming across several levels, and therefore, we see that this effect of enhanced moisture transport in the upward direction is present across several levels. The opposite signed Shapley values for temperature deviations near the surface indicate cold near-surface anomalies are associated with upward moisture fluxes in the boundary layer, as could occur when a CAO occurs over warm ocean waters. The vertical dipole in temperature is indicative of a cold front, where cold air near the surface forces warm moist air to rise. It is worth noting that the sign of the $T-(Q)_{SGF}$ relationship is opposite to the expectation that reduced stability would lead to larger upward fluxes. However, this could indicate that enhanced stability is needed to suppress ordinary upright convection and allow the conditions for CSI.
Since cold air outbreaks cause dry anomalies close to the surface, we can explain the $Q-(T)_{SGF}$ matrix by arguing that this dry anomaly causes stronger surface fluxes, translating to stronger upward heat fluxes throughout the boundary layer. 

The effect of meridional wind on the moisture flux can also be understood as cold, dry air entering the region from the north (negative meridional wind deviation), descending over the warm ocean, and triggering an instability or forcing warmer, moist air to move upward. Therefore, there is a larger upward heat and moisture flux in cases with near-surface northerly anomalies, which is observed both in the $V-(Q)_{SGF}$ and $V-(T)_{SGF}$ matrices. The opposite signed meridional wind anomalies at the top of the boundary layer can be interpreted as a signature of the warm-sector air mass that is being replaced. Similarly, upward fluxes associated with fronts are also evident in change in wind direction evident in the Shapley matrices involving $\partial_xV$ (Figure \ref{Figure: SHAP sign matrices}). The $V-(U)_{SGF}$ matrix appears with opposite sign to the $V-(Q)_{SGF}$ and $V-(T)_{SGF}$ because zonal momentum is largest in the upper troposphere and the triggering of mesoscale vertical motions acts to move momentum towards the surface, reducing the shear. 

The influence of the zonal wind profile on mesoscale fluxes is characterized by a relationship between increased lower-level zonal wind shear and enhanced upward fluxes, which is most evident in the $U-(T)_{SGF}$ and $U-(Q)_{SGF}$ matrices. This can be understood in terms of the MPV discussion in Section 3.1, because this increased vertical shear, in combination with tilted isentropes, would lead to negative $\zeta_{\theta_e}$. As discussed in the previous section, the influence of $\omega$ primarily shows up in terms of a local relationship where the vertical velocity and the mesoscale flux have the same sign.

\section{Discussion and Conclusions}
\label{sec:discusssion}

Our work has trained an offline ML parameterization for predicting midlatitude mesoscale fluxes, which are not resolved or parameterized in conventional coarse-resolution climate models. We show that the ML model is skillful and use ablation experiments and Shapley analysis to identify the source of this skill and to understand the dynamics governing mesoscale fluxes in midlatitudes. Here, we discuss some of the remaining challenges and present our conclusions. 

\subsection{Challenges and Outlook in Improving Model Performance}
\label{subsec:discussion model performance}

The NATLx8 simulation used to train our ML model has enabled a more targeted parameterization by resolving midlatitude mesoscale that are not explicitly parameterized in global climate models, while remaining sufficiently coarse to separate these scales from smaller-scale convective processes. Furthermore, the 30-year simulation length provides good statistics of climatological relationships. However, the simulation has also led to limitations, including potentially limited spatial generalizability of the ML model due to its restriction to the Gulf Stream region, as higher-resolution grid spacing is confined to the North Atlantic. A natural next step would therefore be to expand the model to other regions.

The performance of our ML model for learning realistic mesoscale fluxes can reach $R^2$ scores of between 0.5 and 0.8 for all vertical fluxes considered at the levels on which they are most important. The ANN is thus able to explain between 50\% and 80\% of the variance, depending on the variable and level, with the exception of levels where variance is low. This skill can partially be attributed to the vertical velocity as a predictor, which ablation experiments revealed to contribute significantly to the performance. This is especially prominent in a local neighbor-to-level mapping, where the performance is predominantly attributed to the vertical velocity. This is potentially problematic and might lead to instabilities in an online implementation of the parameterization, because we showed that coarse-grained vertical velocity might not be representative of vertical velocity from a low-resolution setup, such that vertical velocity might be less informative in an online setting within a low-resolution model. However, since the vertical velocity was a relatively less important predictor in the model with profile-to-profile mapping, this issue can be avoided by using a model that includes vertical profile information in the input features and excludes vertical velocity.

The good performance is also attributed to the inclusion of a large number of input features. Other studies, for example, the offline ML parameterizations in \cite{Wang_nonlocal} and \cite{Yuval_OGorman_2021} have shown moderate skill by using a smaller number of features and simpler architectures. Given that our ablation experiments have shown a significant performance drop to R$^2$ scores around 0.2 when using only the fundamental dry variables, we conclude that more features are needed to capture the more complex relationships in more realistic simulations of Earth's atmosphere. We expect that adding the latent and sensible heat flux, in addition to the features already included in our model, could lead to an additional improvement due to its importance for triggering convection, which we showed to be occurring in the clusters of data with the largest errors. Once a performance ceiling is reached, a natural next step is to move from deterministic, conditional mean predictions to stochastic parameterizations that predict a distribution of subgrid fluxes conditioned on the large-scale state (e.g., via ensembles of ANNs; \cite{Mansfield2024,Behrens2025}).

We explored a few approaches that did not lead to improvement of the performance, and therefore are not included in the result section. Motivated by the fact that some extreme values in the precipitation regimes (clusters 0 and 3) are not well captured, we evaluated the performance of a model trained on a dataset where samples containing extremes above a threshold in any of the subgrid fluxes were removed. Excluding these extremes reduces the MSE loss and the variance, but does not improve the coefficient of determination, and for levels highly sensitive to variance, it can even reduce the $R^2$ score. Removing extreme values might be motivated by other reasons, for example if the extreme values in the high-resolution simulations are unphysical.
Therefore, it might be reasonable to exclude them if the ANN is unable to generalize effectively. Another idea was to incorporate memory by including the variables from the previous time step, which would double the number of input features. This approach did not enhanced the performance, and the Shapley value analysis revealed that the additional memory features had no significant impact on the model's prediction. This is somewhat expected, as six-hourly time steps are too large for a strong local correlation that the ANN could learn from. However, such an approach might work better with higher-temporal-resolution training data.
Including previous time steps would allow the ANN to capture dynamics instead of statistical correlations between the inputs and outputs. 
Motivated by the study by \cite{Shap_Beucler}, which argues that transforming variables into climate invariant quantities leads to better generalization in ML prediction and more localized dependence of inputs and outputs, we tried replacing specific humidity with the relative humidity. While this did not improve the ANN performance in our control simulation setting, it may nevertheless prove valuable for generalizability in simulations of a warming climate.

Finally, we have ideas that can be applied in future studies to further improve the performance. Several auxiliary tasks might help improve the model's performance. A possible task could be determining the regime of the sample. Since there is some difference in performance between the ocean and land, another task might be to determine if a sample lies over land or ocean. Alternatively, land fraction, topography and SST could be included as input features. 
An additional auxiliary task might also be included to improve the local mapping. Since in a neighbor-to-level mapping, the levels are mixed up, as different levels correspond to different samples, an auxiliary task that tries to predict the level might help improve performance in the local mapping. Another idea involves transforming the target distribution to a Gaussian distribution with tools like \texttt{QuantileTransformer} from the \texttt{scikit-learn} library. With this approach, it is expected that during training, the use of balanced losses is not needed, avoiding problems due to the pronounced asymmetric tails of output distributions.

\subsection{Physical Insights from Data-Driven Models}
\label{subsec:discussion physical insights}

In addition to exploring ML approaches to improve global climate models, this study also aimed to provide physical insights.
While it is not guaranteed that a ML model will learn accurate underlying physical processes, we found neural network responses that align with physical expectations. 
For example, we see upward moisture fluxes when there are cold northerly winds blowing over a warm ocean surface with warm sub-saturated air at the top of the boundary layer.

Our results show that the ANN performs worse on two rare precipitation regimes that are represented by clusters containing only 8\% and 0.2\% samples of the dataset. While the MSE is larger in these clusters, the sign of the output fluxes in these clusters is still mostly determined correctly, and sign misclassifications happen more often in other clusters where fluxes have smaller magnitudes. We suggest that increasing the number of samples from these rare clusters in the dataset may improve the generalizability of the model to these regimes and help to understand their dynamics.

Explainable AI methods, such as Shapley values and ablation experiments, can help interpret the behavior of ML models, but they reveal correlations, which do not necessarily imply causation. This makes physical interpretation challenging, especially in cases where there is not a strong a priori understanding of how subgrid-scale fluxes are shaped by large-scale conditions.


\subsection{A Preliminary Stability Heuristic: Shapley-Based Local Response Analysis}

\begin{figure}
\centering
    \includegraphics[width=0.8\linewidth]{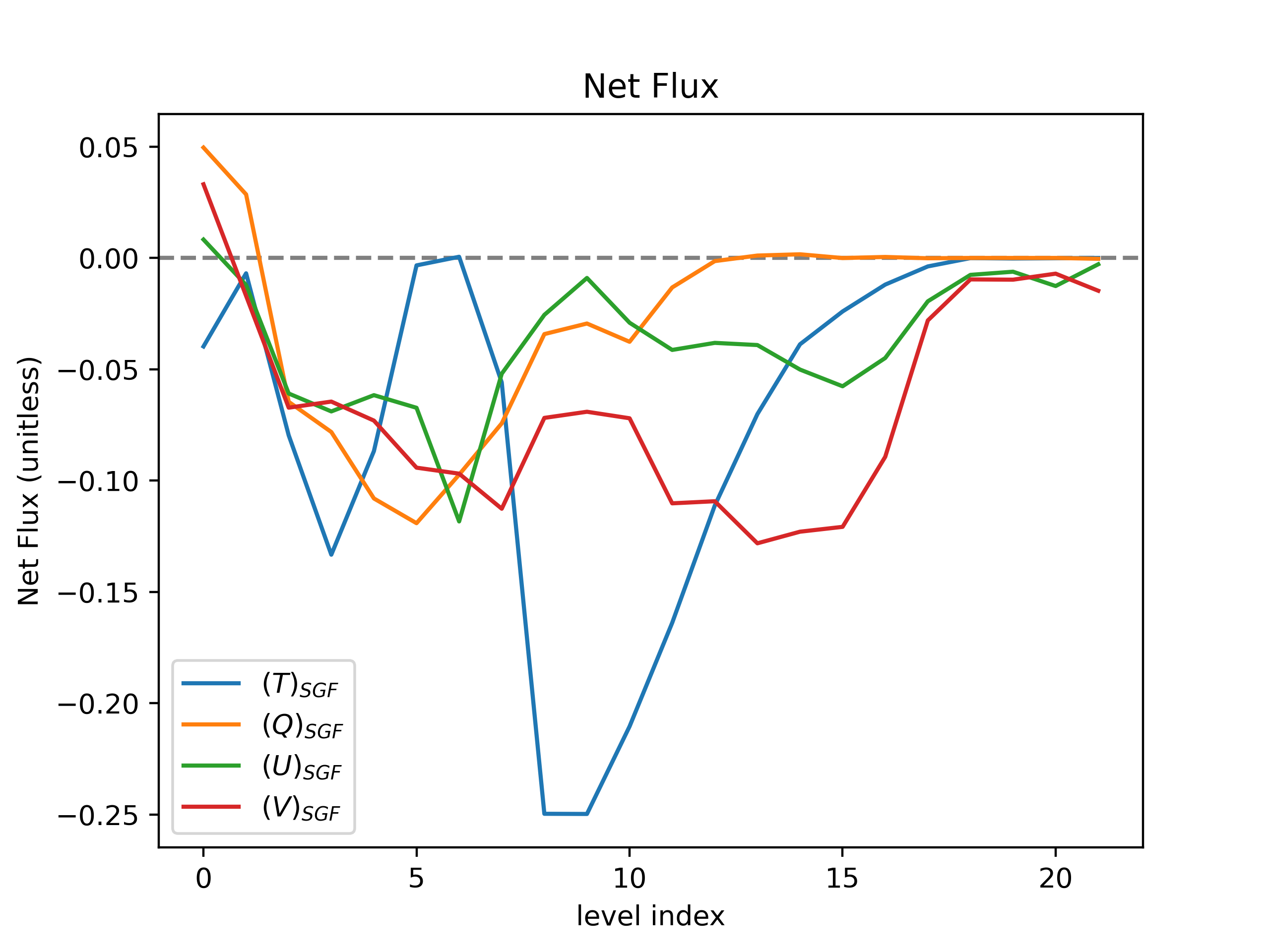}
    \caption{\small Net flux at each level computed from the Shapley values of the ANN (profile-to-profile mapping, excluding vertical velocity as input) for each $(\phi)_{SGF}$ when its tracer $\phi \in $ $\{T,\ Q,\ U,\ V\}$ is increased. The net flux is computed by computing a flux balance involving the flux at a given level index $i \in \{1,..., 22\}$ and the neighboring levels $i-1$ and $i+1$. The level index is increasing for levels closer to the surface.}
    \label{Figure: Net flux}
\end{figure}

Although excluding vertical velocity from the input features might mitigate spurious sensitivities, strong offline performance does not guarantee stable online behavior after coupling the parameterization to the prognostic model. The Shapley matrices can offer a preliminary, local diagnostic of whether instabilities are likely to occur. 
Instabilities are less likely to occur when an increase in a tracer $\phi \in$ \{$U$, $V$, $T$, $Q$\} at a given level $i$ produces a net flux $(\phi)_{SGF}^{\text{net}, i}$ that exports this excess to adjacent levels on average, which is a locally restorative response. To assess this, we evaluate the influence of $\phi$ at a given level $i$ on the flux at the upper level $i+1$, at $i$, and at the lower level $i-1$ using a flux balance approach:
\begin{equation}
    (\phi)_{\text{SGF}}^{\text{net}, i} = -|(\phi)_{SGF}^{i}| + (\phi)_{SGF}^{i+1} - (\phi)_{SGF}^{i-1}
\end{equation}
with outgoing fluxes defined as negative. If the flux of the upper level is directed towards level $i$ the value is positive, therefore this flux is added as an ingoing flux. On the other hand, if the flux of the lower level is directed towards level $i$ the value is negative, therefore this flux is subtracted to represent an incoming flux. For the profile-to-profile mapping ANN excluding the vertical velocity as input, Figure~\ref{Figure: Net flux} shows that the net flux is negative at all levels for temperature, and negative at most levels for momentum (except near the surface). For moisture, positive net flux appears in the two lowest levels and weakly in upper levels where fluxes are close to zero. This suggests that moisture and near-surface momentum responses may be more susceptible to locally amplifying behavior than temperature.

Importantly, this diagnostic is a local, one-step heuristic of locally restorative behavior and should not be interpreted as a formal online stability analysis. It does not account for, e.g., cross-variable couplings, state dependence away from the sampled conditions (including distribution shift), error accumulation over several steps, timestep sensitivity, or feedbacks with the resolved dynamics. A proper online stability assessment requires coupled online integrations or perturbation-growth analyses of the coupled model--parameterization system, as discussed by \cite{brenowitz_interpreting_2020}. If online instabilities emerge, potential mitigation strategies include additive or multiplicate input-noise during training to reduce sensitivity to shifted inputs (\cite{heuer2025trainingdataconfidenceguidedmixing}) and, pragmatically, screening checkpoints, architectures, and training configurations to identify ML parameterizations that are more likely to remain stable online when instability cannot be directly mitigated (\cite{Lin2025}).

\subsection{Conclusions}

In conclusion, our results can be summarized as follows:
\begin{itemize}
    \item Many features, including vertical profile information of the input variables, are needed for skillful prediction of mesoscale fluxes in midlatitudes.
    \item The performance of the ANN in a non-local mapping is moderate to good across all levels with finite variance in the troposphere.
    \item Zonal momentum flux is the most difficult subgrid flux to predict.
    \item Vertical velocity contributes substantially to the skill of the ML models, especially in local mappings. However, its inclusion in an online parameterization is not recommended because the coarse-grained vertical velocity is not representative of the vertical velocity from a low-resolution model. When excluding vertical velocity, the level-to-level mappings perform only slightly better than a level-wise climatological baseline.
    A neighbor-to-level mapping without vertical velocity achieves performance comparable to a level-to-level mapping that includes $\omega$.
    \item Near-surface divergence terms and the vertical profile of temperature are some of the most useful input features for the prediction of mesoscale fluxes with the ANN. 
    \item The ANN is skillful even without the inclusion of moist variables (or vertical velocity). Therefore, a large amount of the mesoscale fluxes in midlatitudes could be explained by dry variables alone.
    \item Moisture variables have a strong non-local influence on subgrid fluxes, especially moisture and heat fluxes.
    \item Cold air out breaks are important: dry, cold anomalies and northerly wind near the surface lead to upward heat and moisture fluxes across the boundary layer. This relationship supports the idea CAOs might be crucial for explaining mesoscale fluxes caused by slantwise convection.
\end{itemize}

\section*{Acknowledgments}

R.C.J.W acknowledges funding from the Swiss National Science Foundation (Award PCEFP2 203376). T.B. acknowledges funding from the Swiss State Secretariat for Education, Research and Innovation (SERI) for the Horizon Europe project AI4PEX (Grant agreement ID: 101137682 and SERI no 23.00546).

\section*{Data Availability}
The coarse-grained input and output data over the study region for training the ML models are processed as described in the text from simulation output from \cite{wills_resolving_2024} and are available on Zenodo (\cite{Zenodo}), along with code for all ML models, their training, and their Shapley analysis. 

\begin{footnotesize}
\printbibliography
\end{footnotesize}

\newpage
\renewcommand{\thefigure}{S\arabic{figure}}
\renewcommand{\thetable}{S\arabic{table}}
\renewcommand{\thesection}{S\arabic{section}}

\setcounter{figure}{0}
\setcounter{table}{0}
\setcounter{section}{0}

\title[SI: Machine Learning of Mesoscale Fluxes in Midlatitudes]{Supporting Information for "Machine Learning of Vertical Fluxes by Unresolved Midlatitude Mesoscale Processes"}

\author{Erisa Ismaili; Robert C. Jnglin Wills; Tom Beucler}

\address{ETH Zürich, University of Lausanne}
\ead{r.jnglinwills@usys.ethz.ch} 
\vspace{10pt}

\section{ANN architecture and Loss Curves}

\begin{table}[h]
    \centering
    \caption{\small Overview of fixed and explored hyperparameters and their respective ranges/ values}
    \begin{tabular}{lll}
        \hline
        \bf Hyperparameter & \bf Explored & \bf Range / Values \\ 
        \hline
        Early stopping criterion & No & Minimum validation MSE \\
        Batch size & No & 1024 \\
        num\_workers & No & 6 \\
        Learning rate scheduler & No & CyclicLR \\
        Batch normalization & No & Enabled \\
        Optimizer & Yes & Adam, RAdam \\ 
        Epochs & Yes & 30 -- 150 \\
        Network width & Yes & $4 \cdot 128$ -- $14 \cdot 128$ (layer-dependent) \\
        Network depth & Yes & 3 -- 18 \\
        Activation function & Yes & ReLU, GELU, ELU, Sigmoid \\
        Dropout rate (excluding GELU) & Yes & 0.005 -- 0.03 \\
        Loss function & Yes & MSE, MAE, MSLE, Huber, \\
        & & Balanced L1, custom functions \\
        Weight decay & Yes & $10^{-7}$ -- $10^{-3}$ \\
        Base learning rate & Yes & $10^{-6}$ -- $10^{-4}$ \\
        Scheduler step size & Yes & $10^{2}$ -- $10^{3}$ \\
        \hline
    \end{tabular}
    \label{tab: hyperparameters}
    
    \vspace{0.3em}
    {\footnotesize
    \textbf{Abbreviations:}
    MSE — Mean Squared Error;
    MAE — Mean Absolute Error;
    MSLE — Mean Squared Logarithmic Error;
    ReLU — Rectified Linear Unit;
    GELU — Gaussian Error Linear Unit;
    ELU — Exponential Linear Unit;
    Adam — Adaptive Moment Estimation;
    RAdam — Rectified Adam.
    }
\end{table}

\begin{table}[h]
    \centering
    \caption{\small Architecture of the neural network, showing the depth and layers.}
    \begin{tabular}{ll}
        \hline
        \bf Depth & \bf Layers \\ 
        \hline
        Layer 1 & Linear(244, 1792), GELU(), BatchNorm(1792) \\ 
        Layer 2 & Linear(1792, 1536), GELU(), BatchNorm(1536) \\ 
        Layer 3 & Linear(1536, 512), GELU(), BatchNorm(512) \\ 
        Layer 4 & Linear(512, 768), GELU(), BatchNorm(768) \\ 
        Layer 5 & Linear(768, 512), GELU(), BatchNorm(512) \\ 
        Layer 6 & Linear(512, 384), GELU(), BatchNorm(384) \\ 
        Layer 7 & Linear(384, 384), GELU(), BatchNorm(384) \\ 
        Layer 8 & Linear(384, 88) \\ 
        \hline
    \end{tabular}
    \label{tab: network layers}
\end{table}

\begin{figure}[H]
\centering
    \includegraphics[width=0.75\linewidth]{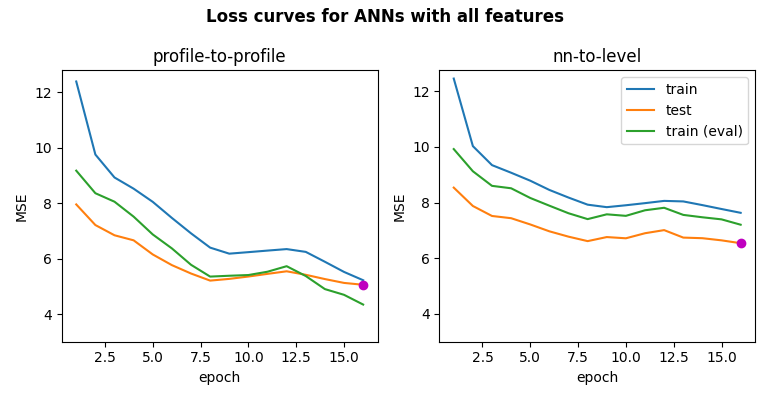}
    \caption{Evolution of the losses on the training and test set while training. The model was trained for 16 epochs. The number of epochs is chosen by early stopping based on the minimum loss on the validation set. This minimum is also the minimum of the test loss (dot in magenta).  While a balanced L1 loss is used during training, the evaluation of the training set (train) and the test set is performed using MSE. The plots show the training loss with the model being in training mode and evaluation (eval) mode. Left are the curves for the profile-to-profile mapping and right are the curves for the neighbor-to-level mapping.}
    \label{fig: loss functions}
\end{figure}

\section{Features of the K-means Clustering and Data Properties}

\begin{table}[H]
\centering
\caption{Overview of all features and diagnostic quantities used in the clustering analysis.}
\begin{tabular}{llp{7.5cm}}
\hline
\textbf{Features} & \textbf{Symbol}  \\
\hline
Vertical zonal wind shear & $U_{500}-U_{850}$\\

Vertical meridional wind shear & $V_{500}-V_{850}$\\

Horizontal divergence at lowest level & $D_{\text{bot}}$ \\

Relative vorticity at 500 hPa & $\zeta_{500}$\\

Temperature difference & $T_{\text{bot}} - T_{850}$\\

Near-surface relative humidity & $\mathrm{RH}_{\text{bot}}$\\

Surface pressure & $P_S$\\

Convective Available Potential Energy & $\mathrm{CAPE}$\\

Subgrid-scale zonal flux at 500 hPa & $(U)_{\text{SGF, }500}$ \\

Subgrid-scale meridional flux at 500 hPa & $(V)_{\text{SGF, }500}$ \\

Subgrid-scale temperature flux at 500 hPa & $(T)_{\text{SGF, }500}$ \\

Subgrid-scale moisture flux at 500 hPa & $(Q)_{\text{SGF, }500}$  \\
\hline
\\
\hline
\textbf{Additional Diagnostic Variables} & \textbf{Symbol}  \\
\hline
Meridional wind at the lowest level & $V_{\text{bot}}$\\
Precipitation resulting from large-scale motion & PRECL\\

Precipitation resulting from parameterization of convection & PRECC\\

Equivalent potential temperature difference & $\theta_{e, 500}-\theta_{e, 850}$\\

\hline
\end{tabular}
\label{tab: cluster features and vars}

\end{table}

\begin{table}[H]
\centering
\caption{\small Definition and characterization of the clusters}
\begin{tabular}{p{1.3cm} p{5.5cm} p{2.3cm} p{2cm}p{1.3cm} } 
 \hline
 \bf Cluster & \bf Description & \bf Regime-definition & \bf CSI relevance & \bf Per-centage \\ 
\hline
 \bf 0 & Similar to cluster 3 but weaker characteristics & Precipitation and convection & high & 8\% \\ 
 \hline
 \bf 1 & High pressure, mainly low values of subgrid-scale fluxes  & Stable & low & 45\% \\ 
 \hline 
 \bf 2 & Large wind shears, high vorticity, large temperature differences, dry & CAO & very high & 24\% \\ 
 \hline
 \bf 3  & High convergence, inclusion of low temperature differences and cyclones, contains highest subgrid-scale fluxes for each variable, high CAPE & Heavy precipitation & high & 0.3\% \\ 
 \hline
 \bf 4 & Low pressure, low temperature differences, moist & Warm moist air masses & high & 23\% \\ 
\hline
\end{tabular}
\label{tab: cluster characterization}
\end{table}

\begin{figure}[H]
    \centering
    \includegraphics[width=1\linewidth]{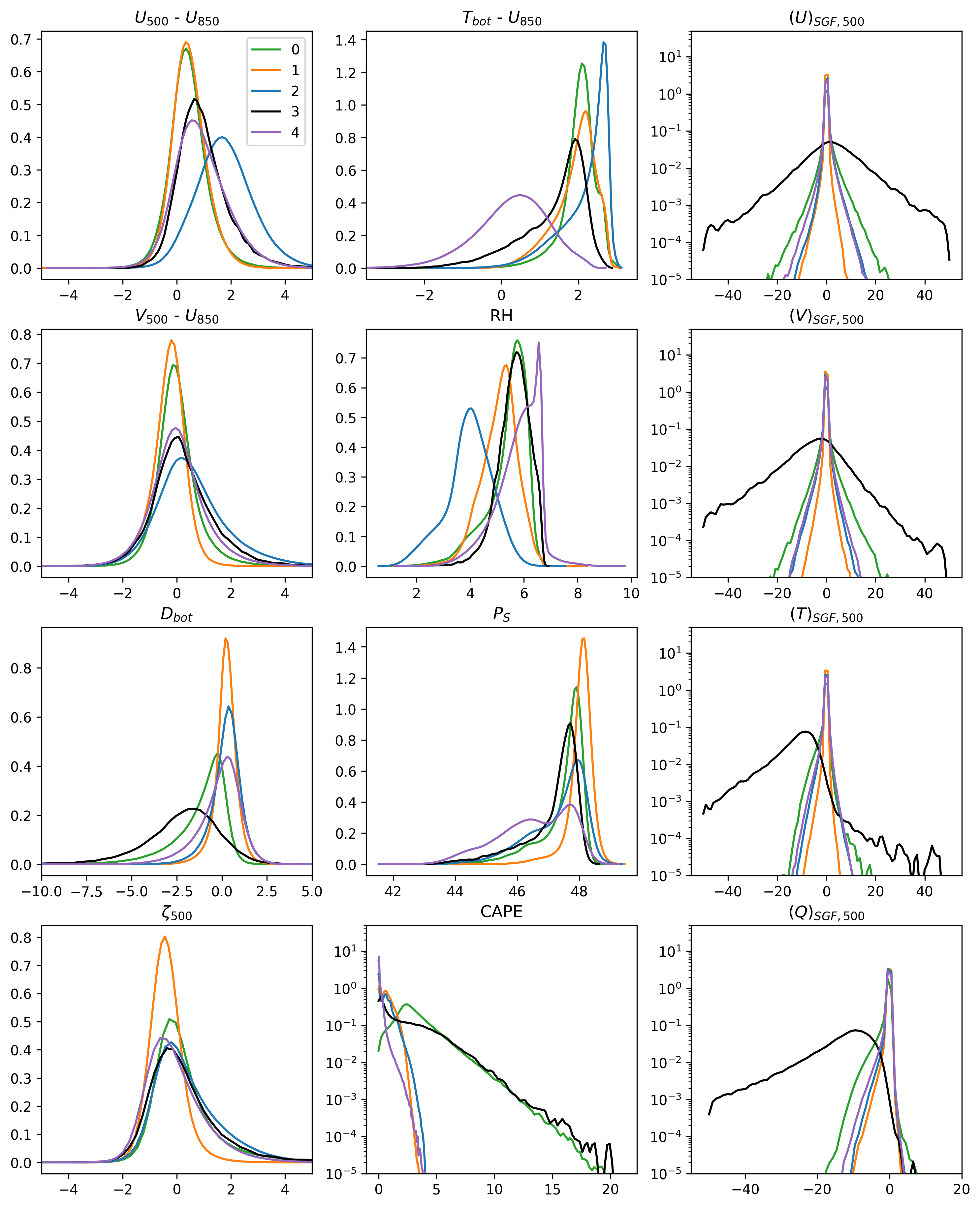}
    \caption{Kernel density estimations for the cluster distribution of all features used in the K-means++ clustering algorithm.
    }
    \label{fig:distributions_kmeans_all}
\end{figure}

\begin{table}[H]
\centering
\caption{Statistical properties and positive-value proportions of subgrid-scale fluxes nearest to 500 hPa separated by cluster. The units of the mean and median are according to the normalization used for the ANN training and, therefore, arbitrary and comparable across the different fluxes.}
\label{tab:cluster-stats-arbitrary}
\begin{tabular}{l c c c c c}
\toprule
\textbf{Variable} & \textbf{Cluster} & \textbf{Mean} & \textbf{Median} & \textbf{Skewness} & \textbf{Positive (\%)} \\
\midrule
\multirow{5}{*}{$(U)_{\text{SGF}}$}
 & 0 & 0.72 & 0.059 & -1.1 & 55 \\
 & 1 & 0.006 & 0.007 & -190 & 52 \\
 & 2 & 0.52 & 0.11 & 6.3 & 60 \\
 & 3 & 14 & 11 & 1.6 & 61 \\
 & 4 & 0.28 & 0.049 & -4.3 & 56 \\
\midrule
\multirow{5}{*}{$(V)_{\text{SGF}}$}
 & 0 & -0.27 & -0.051 & -2.6 & 45 \\
 & 1 & -0.08 & -0.018 & -26 & 45 \\
 & 2 & -0.36 & -0.086 & -15 & 41 \\
 & 3 & -30 & -21 & -6.9 & 29 \\
 & 4 & -0.26 & -0.054 & -2.1 & 43 \\
\midrule
\multirow{5}{*}{$(Q)_{\text{SGF}}$}
 & 0 & -2.2 & -0.24 & -4.6 & 28 \\
 & 1 & -0.13 & -0.003 & -15 & 48 \\
 & 2 & -0.24 & -0.006 & -11 & 45 \\
 & 3 & -69 & -57 & -2.4 & 0.1 \\
 & 4 & -0.46 & -0.012 & -9.2 & 42 \\
\midrule
\multirow{5}{*}{$(T)_{\text{SGF}}$}
 & 0 & -1.3 & -0.033 & -1.3 & 47 \\
 & 1 & -0.052 & -0.002 & -6.5 & 49 \\
 & 2 & -0.24 & -0.014 & -0.91 & 49 \\
 & 3 & -75 & -62 & -2.1 & 1.0 \\
 & 4 & -0.28 & 0.000 & -3.9 & 50 \\
\bottomrule
\end{tabular}
\label{tab:stat by cluster}
\end{table}

\section{ANN Performance: Analysis by Region and by Cluster}

\begin{figure}[H]
\centering
    \includegraphics[width=0.8\linewidth]{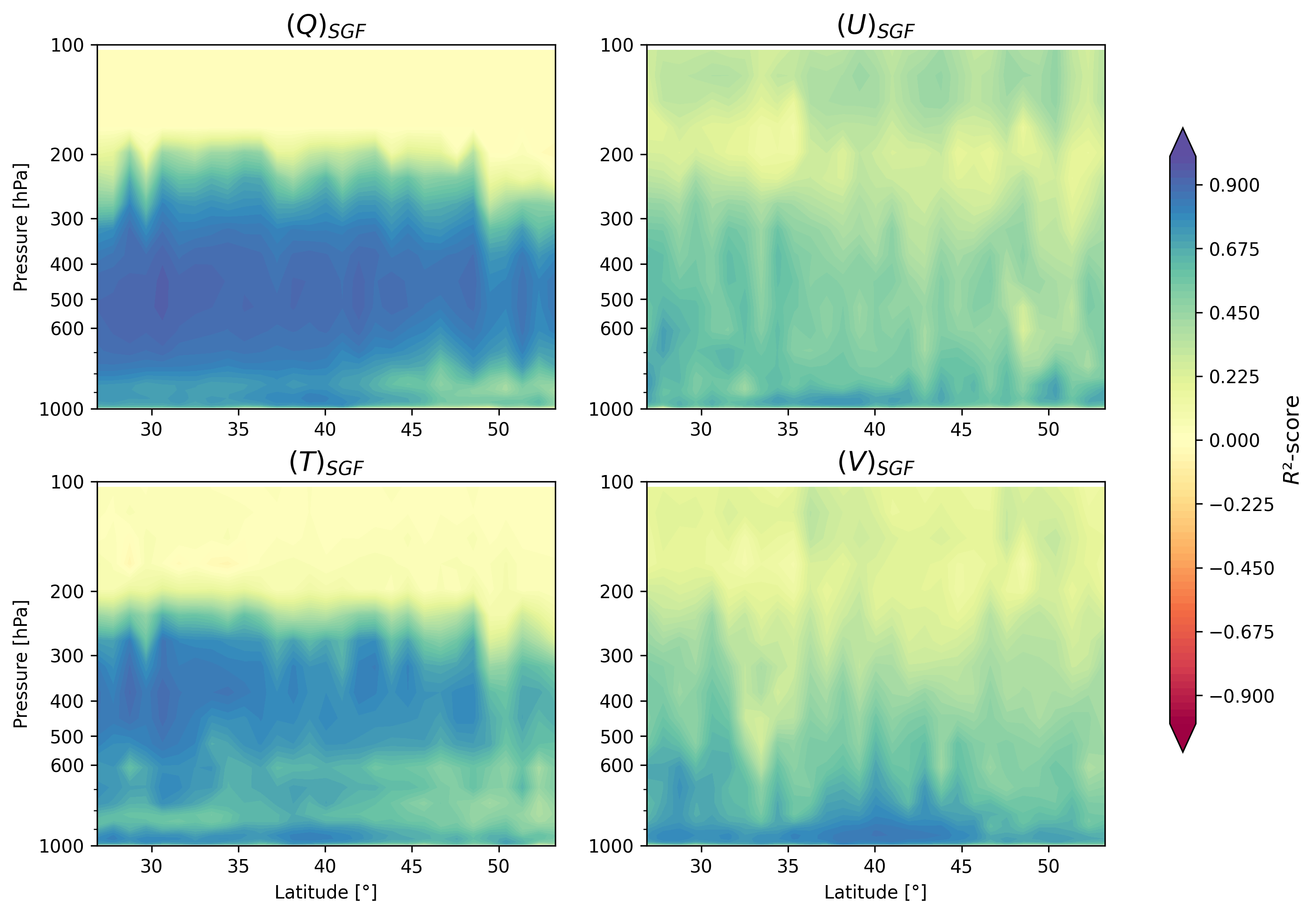}
    \caption{Zonal mean $R^2$ scores of the predicted mesoscale fluxes.}
    \label{fig: r2 altitude latitude}
\end{figure}
\begin{figure}[H]
\centering
    \includegraphics[width=1.0\linewidth]{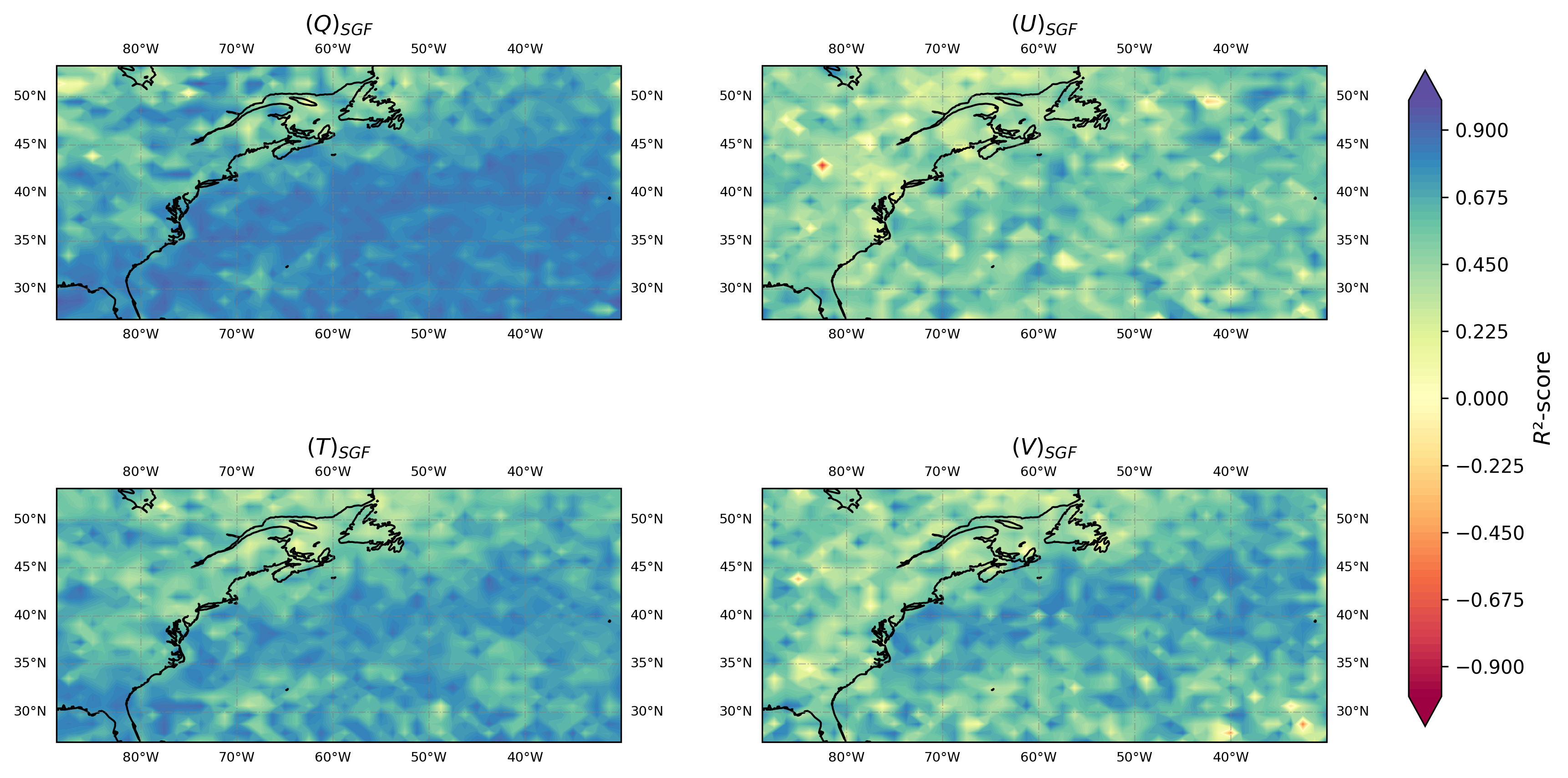}
    \caption{Vertically averaged $R^2$ scores of the predicted mesoscale fluxes on the map.}
    \label{fig: r2 map}
\end{figure}

\section{Shapley Value Analysis}

\begin{figure}[H]
    \centering
    \includegraphics[width=0.65\linewidth]{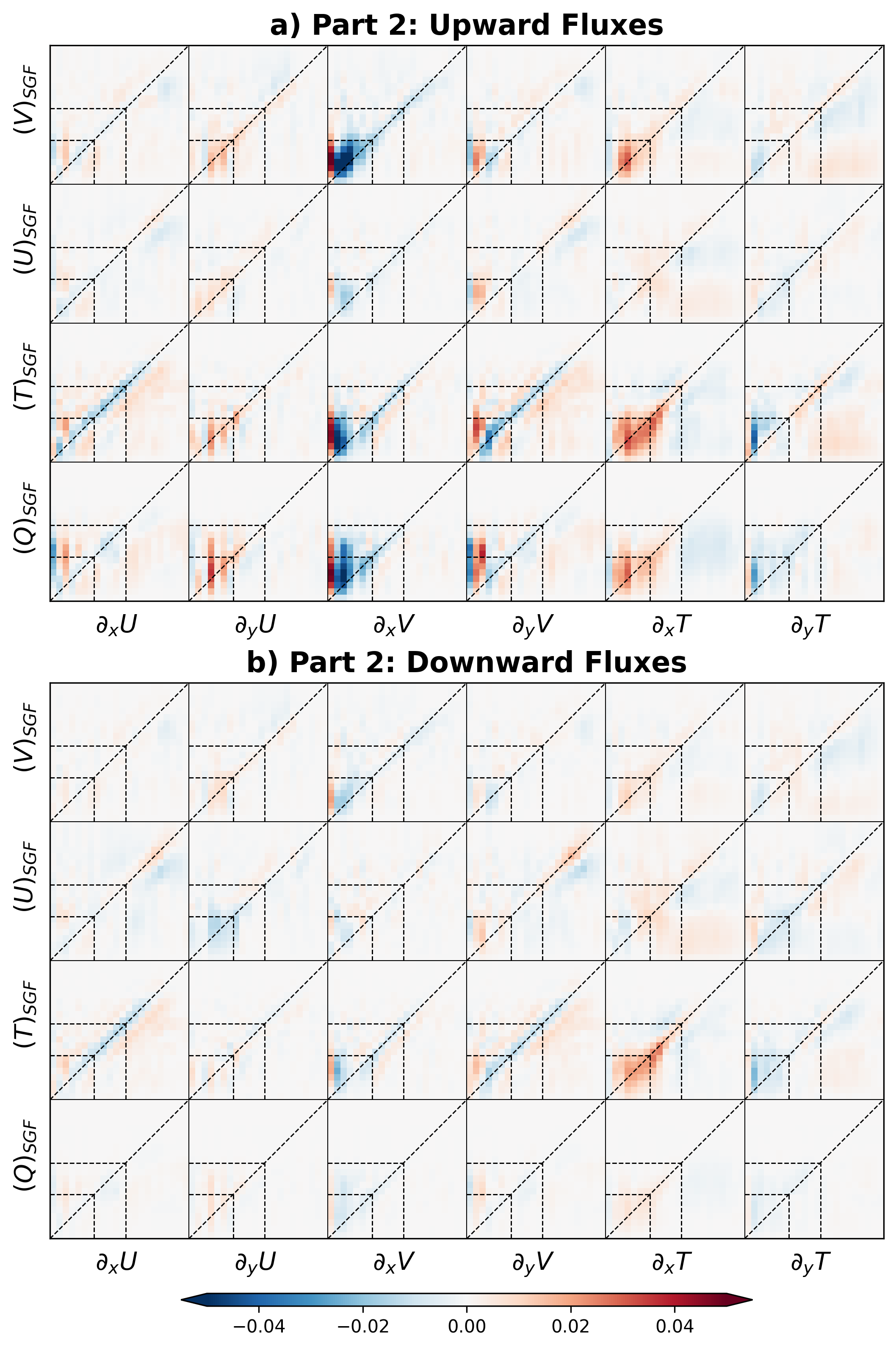}
    \caption{\small Part 2 of Shapley responses separated by upward and downward fluxes, showing Shapley response matrices for inputs involving horizontal derivatives of the state variables.}
    \label{Figure: SHAP sign matrices}
\end{figure}

\end{document}